\documentclass[journal]{IEEEtran}

\ifCLASSINFOpdf
\usepackage[pdftex]{graphicx}
\usepackage{graphicx}

\usepackage[left=0.75in,right=0.75in,top=1in,bottom=1in]{geometry}

\usepackage{subfigure}

\usepackage{lmodern}
\usepackage[T1]{fontenc}

\usepackage{latexsym}
\usepackage{amssymb}
\usepackage{amsmath} 
\usepackage{enumerate}
\usepackage{mathtools}
\usepackage{hyperref}
\usepackage{multirow}
\usepackage{makecell}
\usepackage{longtable} % for 'longtable' environment
\usepackage{pdflscape,array,booktabs}
\usepackage[table]{xcolor}
\usepackage{footnote}
\usepackage{hhline}
\usepackage{float}

\usepackage{cite}

\usepackage{enumitem}
\setlist[itemize]{noitemsep, topsep=0pt, leftmargin=*}

\usepackage[font=small,skip=2pt]{caption}

\usepackage[ruled, vlined, linesnumbered]{algorithm2e}

\def\ShadingColor{0.95}

\definecolor{darkcerulean}{rgb}{0.03, 0.27, 0.49}
\definecolor{darkcandyapplered}{rgb}{0.64, 0.0, 0.0}
\definecolor{darkspringgreen}{rgb}{0.09, 0.45, 0.27}
\definecolor{lightgray}{rgb}{0.83, 0.83, 0.83}

\usepackage{xcolor}

\def\RevisionColor{black!50!black}

\usepackage[switch,pagewise]{lineno}

\definecolor{brightmaroon}{rgb}{0.76, 0.13, 0.28}

\usepackage{flushend}

\ifCLASSINFOpdf
\else
\fi
\begin{document}

\title{\textbf{\huge EMPIOT: An Energy Measurement Platform\\ for Wireless IoT Devices}}

%% **************************************************

% \author{Behnam~Dezfouli,~\IEEEmembership{Member,~IEEE},
%         Immanuel~Amirtharaj,~\IEEEmembership{Student Member,~IEEE,}\\
%         Chia-Chi~Li,~\IEEEmembership{Student Member,~IEEE}% <-this % stops a space

% \thanks{ 
% 	% Please, write here acknowledgment for financial support if desired.
% 	%Cognitive work was supported in part by the Spanish Council of Research under Grant  BS123456 (sponsor and financial support acknowledgment goes here). Paper titles should be written in uppercase and lowercase letters, not all uppercase. Avoid writing long formulas with subscripts in the title; short formulas that identify the elements are fine (e.g., "Nd-Fe-B"). Full names of authors are preferred in the author field. 
% }
% \thanks{Behnam Dezfouli, Immanuel Amirtharaj, and Chia-Chi Li, are with the Internet of Things Research Lab (IoTLAB), Department of Computer Engineering, Santa Clara University, Santa Clara, CA, 95053 USA.
% 	E-mail: \{bdezfouli, iamirtharaj, cli1\}@scu.edu.
% }% <-this % stops a space
% }

%% **************************************************

\author{\IEEEauthorblockN{Behnam Dezfouli\IEEEauthorrefmark{1},
Immanuel Amirtharaj\IEEEauthorrefmark{2}, and Chia-Chi (Chelsey) Li\IEEEauthorrefmark{3}}\\
\IEEEauthorblockA{\IEEEauthorrefmark{1}\IEEEauthorrefmark{2}\IEEEauthorrefmark{3}Internet of Things Research Lab, Department of Computer Engineering, Santa Clara University, USA
\\
    \IEEEauthorblockA{\IEEEauthorrefmark{3}Intel Corporation, Santa Clara, USA}
    \\
\IEEEauthorrefmark{1}\texttt{bdezfouli@scu.edu,}
\IEEEauthorrefmark{2}\texttt{iamirtharaj@scu.ed}u,
\IEEEauthorrefmark{3}\texttt{chelsey.li@intel.com}
}}


\markboth{Journal of Network and Computer Applications, Volume 121, 1 November 2018, Pages 135-148}%
{Dezfouli \MakeLowercase{\textit{et al.}}: EMPIOT: An Energy Measurement Platform \\ for Wireless IoT Devices}



\maketitle

{\color{\RevisionColor}
\begin{abstract}
Profiling and minimizing the energy consumption of resource-constrained devices is an essential step towards employing IoT in various application domains.
Due to the large size and high cost of commercial energy measurement platforms, alternative solutions have been proposed by the research community.
However, the three main shortcomings of existing tools are complexity, limited measurement range, and low accuracy.
Specifically, these tools are not suitable for the energy measurement of new IoT devices such as those supporting the 802.11 technology.
In this paper we propose EMPIOT, an accurate, low-cost, easy to build, and flexible power measurement platform.
We present the hardware and software components of this platform and study the effect of various design parameters on accuracy and overhead.
In particular, we analyze the effects of driver, bus speed, input voltage, and buffering mechanism on sampling rate, measurement accuracy and processing demand.
These extensive experimental studies enable us to configure the system in order to achieve its highest performance.
We also propose a novel calibration technique and report the calibration parameters under various settings.
Using five different IoT devices performing four types of workloads, we evaluate the performance of EMPIOT against the ground truth obtained from a high-accuracy industrial-grade power measurement tool.
Our results show that, for very low-power devices that utilize 802.15.4 wireless standard, the measurement error is less than 3.5\%.
In addition, for 802.11-based devices that generate short and high power spikes, the error is less than 2.5\%.
\end{abstract}
}

\begin{IEEEkeywords}
Energy Efficiency, Testbed, Accuracy, Linux, I2C Driver, 802.15.4, 802.11.
\end{IEEEkeywords}

% \fi
\IEEEpeerreviewmaketitle

\section{Introduction}
\label{intro}
\IEEEPARstart{T}{he} importance of low-cost and accurate energy measurement of IoT devices is justified by two important observations: 
% importance and the effect of the energy consumption of Internet and its connected devices has been recently studied \cite{Baliga2009,Hinton2011,Chiaraviglio2012}.
First, the percentage of energy consumed by connected devices is increasing due to the significant growth in the number of IoT devices.
Gartner predicts the number IoT connected devices will surpass 20 billion by 2020 \cite{Gartner_IoT}.
Second, IoT devices are mostly battery-powered or rely on energy harvesting.
Therefore, it is important to measure, profile, analyze, and improve the energy efficiency of these devices to satisfy the QoS requirements of applications.
% To this end, the research community has been proposed mechanisms for medium access control, transmission power control, rate adaptation, and scheduling.

%Recent studies show the inaccuracy of using simulation for energy profiling \cite{Kaup2014,Gomez2012a,Zhou2013a,Trathnigg2008,eriksson2009accurate}.
%Simulation tools rely on the energy consumption of hardware components in different operational modes.
%However, the energy consumption values reported in dustsheets are either not comprehensive, or they only provide an estimation.
%More importantly, it is hard to predict the actual behaviour of hardware in real-world conditions.
%For example, packet loss and contention for medium access is hard to predict.
%Therefore, reliable and real-time power profiling requires in-situ monitoring of energy consumption.
%Nemo: A high-fidelity noninvasive power meter system for wireless sensor networks-
% this works mentioned that in one wsn deployment some nodes died due to no reason. 
%so we need real-time monitoring of nodes

%TODO: mention here that the existing power measurement ics are not suitable for accurate and high speed measurements.
%refer to the actual IC names and highlight the drawbacks
%this is similar to the way Dutta shows in iCount that the existing battery monitoring systems ADC accuract is low.

{\color{\RevisionColor}
Analytical (and simulation-based) energy estimation tools multiply the time spent in each state (e.g., sleep, processing, transmission/reception) by the power consumed in that state.
This approach, however, is not accurate due to the following reasons \cite{Kaup2014,Gomez2012a,Zhou2013a,Trathnigg2008,eriksson2009accurate}:
(i) Most of the proposed models focus on simple wireless technologies such as 802.15.4 and LoRa \cite{wang2006realistic,martinez2015power}.
However, as new technologies such as 802.11 and LTE are being adopted by IoT, it is important to profile the energy efficiency of devices using these complex technologies.
Specifically, it is difficult to simulate and evaluate the properties of real-world environments (e.g., interference), physical layer, and MAC layer parameters, and study their effects on energy consumption \cite{Li2013b,dezfouli2015modeling,dezfouli2014cama}.
For example, while the energy consumption characteristics of 802.15.4 devices are mostly affected by MAC parameters, 802.11's physical layer parameters are diverse and significantly affect energy consumption. 
In addition, 802.11 offers sophisticated MAC mechanisms (such as access categories, frame aggregation, automatic power save delivery) where predicting their effect on energy consumption is considerably more difficult compared to 802.15.4's MAC.
To show this complexity, for example, our testbed results reported in Figure \ref{fig:ecdf_power} illustrate the significant effect of 802.11 physical layer parameters on current consumption.
The high variations of power consumption versus physical layer parameters justify the importance of extensive empirical measurements under various configurations in order to model and profile the energy characteristics of 802.11 devices.
%
%
% \begin{table}[!t]
% \centering
% \caption{Energy consumption of transmitting UDP packets during one second using an 802.11-based IoT device (CYW943907 \cite{CYW43907AEVAL}).}
% \label{cyw_exp_udp}
% \begin{tabular}{|c|c|}
% \hline
% Configuration        &    Energy (J)     \\ \hline\hline
% AC\_BK, Two Antenna, w/o Interference         &    1.93    \\ \hline
% AC\_BK, Two Antenna, w/ Interference          &     1.31    \\ \hline
% AC\_BK, Single Antenna, w/o Interference         &    1.95     \\ \hline
% AC\_BK, Single Antenna, w Interference       &    1.32     \\ \hline
% AC\_VO, Two Antenna, w/o Interference         &    1.69     \\ \hline
% AC\_VO, Two Antenna, w/ Interference       &    1.5     \\ \hline
% \end{tabular}
% \end{table}
%
%
\begin{figure}[!t]
\centering
\includegraphics[width=0.95\linewidth]{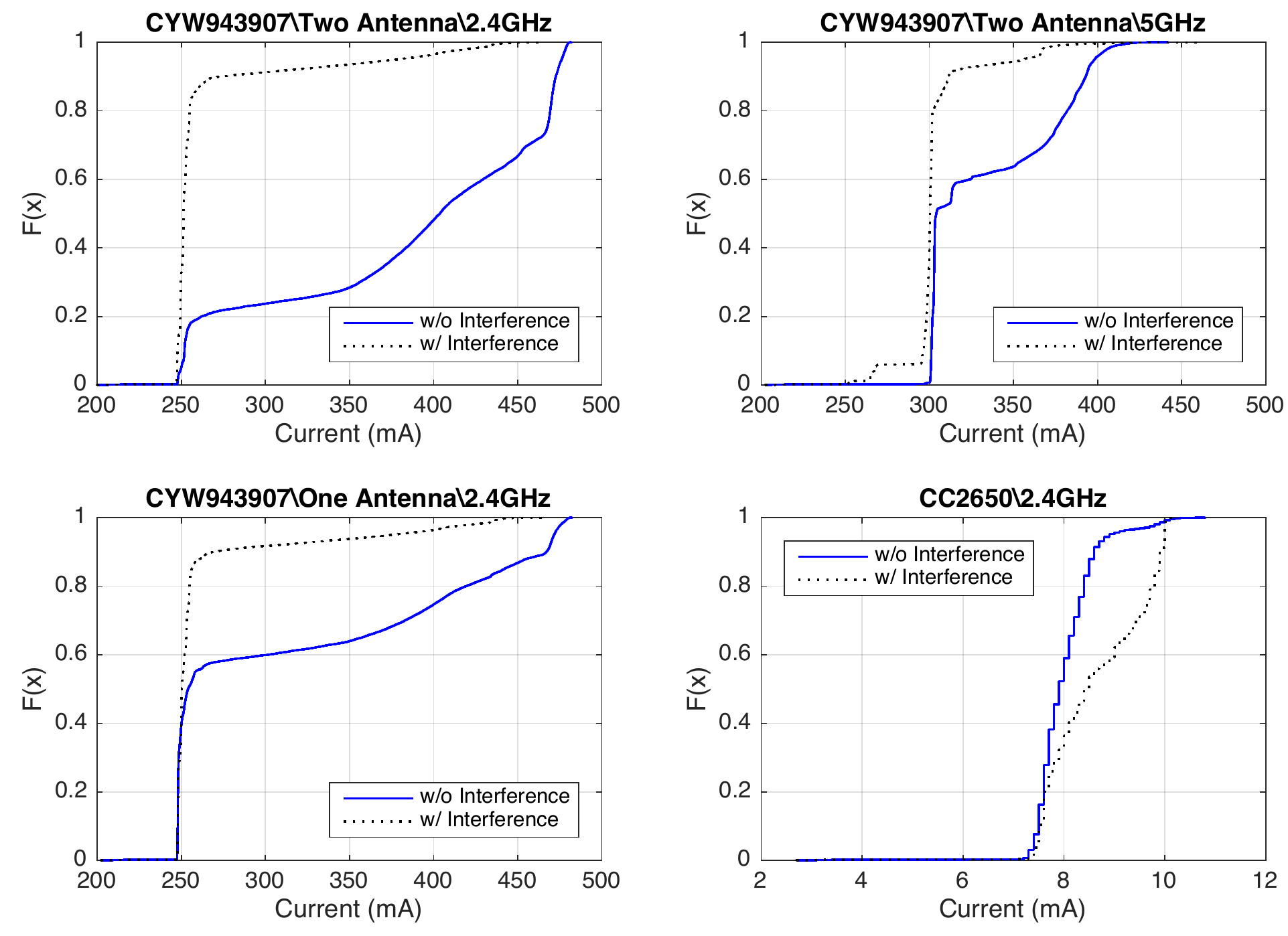}
\caption{Empirical CDF of current consumption for an 802.11-based IoT device (CYW943907 \cite{CYW43907AEVAL}) versus an 802.15.4 IoT device (CC2650 \cite{CC2650}).}
\label{fig:ecdf_power}
\end{figure}
(ii) The energy consumption of all the operational modes of a given system-on-chip (SoC) may not be available.
For example, the datasheet may include only the average current consumption for nominal transmission power values.
As another example, the energy consumption of processors depends on utilization level, frequency scaling, and I/O operations \cite{Stathopoulos2008}.
(iii) IoT boards usually include a SoC and several peripheral components (e.g., ADC, sensors, memory).
Therefore, even if the energy characteristics of the SoC are known, estimating the total energy consumption of the board is very challenging.
(iv) Code structure, algorithms, and data structures, affect the energy consumption of the IoT system.
For example, when a cache memory is present, the type of the data structures used affects cache performance when memory allocation is contiguous.
(v) Analytical models usually do not take into account the cost of start-up energy.
However, the start-up energy cost of a sensor that is periodically woken up to collect a sample can significantly contribute to the total energy cost \cite{brusey2016energy}. 
Despite these shortcomings, most of the research contributions on low-power wireless IoT systems rely on analytical and simulation-based energy estimation due to their simplicity.
}

Although the commercial energy measurement platforms provide very high accuracy, they are costly and bulky.
For example, Keysight 34465A \cite{34465A}, which has 2MB storage and a maximum sampling rate 50Ksps, costs more than \$1300.
Similarly, Keithley 7510 \cite{kth_dmm_7510}, which has 2MB storage and a sampling rate 1Msps, costs more that \$3500.
In addition, due to their size and cost, these devices are not viable solutions to monitor the energy consumption of a large number of IoT devices in a testbed.
Furthermore, their programmability is very limited and inflexible.
These shortcomings also apply to the Monsoon power meter \cite{Monsoon} (cost $\approx$\$800), which is widely-used by academia \cite{manweiler2011avoiding,serrano2015per}.

{\color{\RevisionColor}
Due to these challenges, the research community has proposed several power measurement platforms.
Based on their main shortcoming, these platforms are classified into the following categories:
(i) \textit{complex} (e.g., \cite{Jiang2007,Stathopoulos2008,Duttac2008,Andersen2009,Zhou2013a,Naderiparizi2016}), 
(ii) \textit{limited supported range} (e.g., \cite{Haratcherev2008,Trathnigg2008,Zhu2013b,Hartung2016,Jiang2007,Potsch2017,Potsch2014}),
and (iii) \textit{low accuracy} (e.g., \cite{Gomez2012a,Keranidis2014b}).
The platforms of the first category present complex circuity, which makes the device costly and hard to build.
For the second category, the power measurement platforms cannot be used with a wide variety of IoT devices due to their limited current measurement range, which is a few tens of milliamps.
Specifically, the new generation of IoT devices utilizes wireless technologies (such as 802.11) that result in current spikes as high as 700mA.
For example, Figure \ref{fig:ecdf_power} shows that the current consumption range of the 802.11 device is 46 times wider than that of the 802.15.4 device.
Finally, if the accuracy of an energy measurement platform is low, then it cannot be used for effective study, development and debugging of IoT devices.
In addition to these shortcomings, most of these platforms ignore the effect of voltage variation on energy measurement \cite{Haratcherev2008,Andersen2009,Zhou2013a,Trathnigg2008}.
Furthermore, in terms of accuracy analysis, evaluations are very limited and mostly include only one IoT device type \cite{Jiang2007,Duttac2008,Andersen2009,Zhou2013a}.
}

{\color{\RevisionColor}
In this paper we introduce EMPIOT, an accurate, low-cost, easy to build, and flexible power measurement platform.
EMPIOT has two main components: a shield board, which includes a low-cost INA219 \cite{INA219} energy monitoring chip, and a base board, which runs the controlling and data collection software.
The shield board supports both current and voltage measurement within the operational range of various IoT devices.
% In addition to this shield board, we show how an off-the-shelf INA219 board can be used to build an EMPIOT platform.
% Depending on the configuration used, INA219 can measure currents and voltages as high as 3.2A and 32V, respectively.
The software, written in C++, is composed of two threads for communication with the shield as well as computing energy and saving the data to a file.
Despite the simplicity of its hardware, in this paper we show the effect of various design parameters on performance through extensive experimental studies.
Specifically, we use two I2C drivers, namely BCM \cite{bcmdriver} and Linux driver \cite{linuxi2c}, and show that the BCM driver achieves higher speed, lower energy consumption, and predictable timing characteristics.
We also evaluate the effect of input voltage on sampling rate and measurement accuracy. 
Our results show that reducing the shield's operational voltage increases the conversion time and lowers the sampling rate.
In addition, we confirm that this platform does not require a very reliable power source to achieve high accuracy.
In fact, the accuracy of current and voltage measurement is always within 100$\mu$A and 4mV, respectively.
Based on these results, for those IoT devices with sleep current consumption of less than 100$\mu$A, we propose a hybrid energy measurement model that combines empirical evaluation with analytical modeling.
We also study the effect of batched and continuous file writes, as well as the driver, on the energy consumption of EMPIOT. 
Through empirical evaluations and formulating a mathematical model, we show how the buffering mechanism used affects the power consumption of EMPIOT based on the number of processor cores.

In order to calibrate the platform, we have designed a \textit{programmable calibration tool}, which enables us to generate currents and voltages in a wide range and precisely control the duration of each change.
Using this tool, we calibrate EMPIOT for currents up to 800mA.
We study the accuracy of this platform using five different IoT devices and four types of loads.
Our ground truth is a high-accuracy industrial-grade power measurement tool \cite{kth_dmm_7510}.
Our results confirm that the measurement error is less than 3.5\% for very low-power devices that use the 802.15.4 wireless standard and generate a peak current of 30mA.
In addition, the error is less than 2.5\% for 802.11 devices when their current consumption surpasses 100mA.
We also show that neglecting voltage variations in the energy measurement process may result in up to a 0.5\% increase in measurement error, especially for battery-powered 802.11-based IoT devices.
}

We emphasize that this paper does not intend to propose a complex power measurement tool.
Rather, we study if and how the EMPIOT shield, or even an existing off-the-shelf INA219 breakout board (such as \cite{ina219adafruit}), can be used for accurate power measurement of a wide variety of IoT devices.

The rest of this paper is organized as follows:
Section \ref{platfDesign} presents the platform components and studies the effect of various parameters on design.
The calibration methodology and results are presented in Section \ref{calib}.
Section \ref{perf_eval} studies the accuracy of EMPIOT.
We present related work in Section \ref{related_work}.
We conclude the paper in Section \ref{conclusion}.

\section{Platform Design}
\label{platfDesign}

In this section we present the hardware and software design of EMPIOT.
We also study the implications of design choices on performance through extensive empirical evaluations.

%HINTS: use military/industry grade components

%We observed that CYW43907 and BCM4336 continue their operation even when the voltage drops to as low as 2.7 v and 3.6 v, respectively.

\begin{figure}[!t]
\centering
\includegraphics[width=0.8\linewidth]{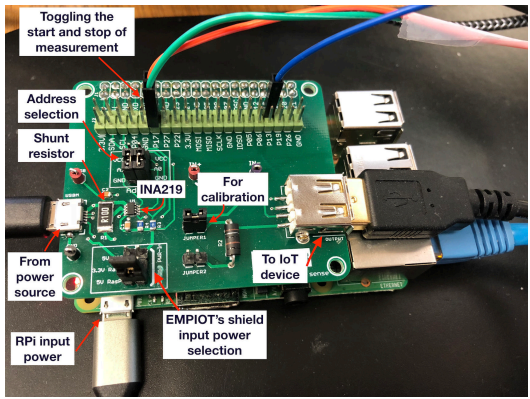}
\caption{Components of EMPIOT.}
\label{fig:empiot_hw}
\end{figure}

\subsection{Hardware}
\label{hw}
{\color{\RevisionColor}
EMPIOT is composed of a shield that is installed on and communicates with a Raspberry Pi (RPi), as shown in Figure \ref{fig:empiot_hw}.
The core of the shield is an INA219 \cite{INA219}, which is a low-cost ($\approx$\$2) current and bus voltage monitoring chip.
INA219 is available in two classes, INA219A and INA219B. 
We use INA219B due to its lower (0.5\%) variations versus temperature.
The energy draw of the chip is 1mA, and it can operate using a 3 to 5.5V supply.
INA219 measures bus voltage directly, and current is measured through digitizing the voltage across a shunt resistor.
In other words, given a shunt value $R_{shunt}$, we can compute current using $i=v_{shunt}/R_{shunt}$, where $v_{shunt}$ is the digitized shunt voltage value.
}

% The ADC's basic resolution is 12 bits, and 1 LSB step size for shunt voltage and bus voltage are 10$\mu$V and 4mV, respectively.
The ADC is a delta-sigma type and uses high frequency (500KHz) to collect analog samples.
After collecting samples, the ADC uses a low-pass digital filter for noise reduction, and a decimator averages the analog samples.
The conversion ready bit is set automatically when a new sample is ready.
The INA219 datasheet mentions that the duration of these operations for 12 and 9-bit sampling resolutions are 532-586$\mu$s, and 84-93$\mu$s, respectively.
However, as we will show in Section \ref{eff_sampl_rate}, the actual sample preparation time is longer than these reported values.

{\color{\RevisionColor}
The full scale voltage range supported across the shunt resistor is 40mV.
Since the ADC resolution is 12 bits, one LSB size for shunt voltage is $\sigma^{shunt}_{v} = 40/(2^{12}-1) \approx 10\mu$V. 
We have used $R_{shunt}=0.1\Omega$ with 0.5\% accuracy.
Therefore, EMPIOT's current resolution is $\sigma_{i}^{shunt}= \sigma^{shunt}_{v}/ R_{shunt} = 10\mu V/0.1\Omega = 100\mu$A for currents up to 400mA.
Depending on the maximum possible current, the power gain amplifier (PGA) can be configured to achieve the full-scale range through dividing shunt voltage by 2, 4, or 8, before digitization.
Therefore, shunt voltage in four various ranges can be measured: $[0, 40]$mV, $[0, 80]$mV, $[0, 160]$mV, $[0, 320]$mV.
The maximum supported bus voltage is either 16V or 32V, depending on the configuration applied.
When configured for 16V, the accuracy of bus voltage measurement is $\sigma_{v}^{bus} = 16/(2^{12}-1) \approx 4$mV. 
Considering the nominal operational range of IoT devices, in this paper we assume current and voltage are less than 800mA and 5.5V, respectively.
}

The shield board communicates with the base board using Inter-Integrated Circuit (I2C) bus.
The minimum and maximum bus speeds supported by INA219 are 0.1MHz and 2.5MHz, respectively.
We chose RPi as the base board to configure the chip and collect the results because: 
(i) its I2C rate is fast enough to support the sample generation rate of INA219,
(ii) the memory card enables power sampling for very long durations,
(iii) it can be used for programming and debugging of the attached IoT device,
and (iv) the existence of multiple communication technologies (Ethernet, WiFi) simplifies testbed setup and remote access.

% This chip provides a sample averaging feature through which the user can read the average of the samples.
% This would also be helpful when the deign does not want to involve the processor in the averaging process.
% For example, when using the 12-bit mode with 128 sample averaging, the processor can read the value every $532\mu$s $\times 128 = 68$ms.

\subsection{Software}
The pseudo-code of the software is given in Algorithm \ref{alg:empiot_sw}.
At a high level, the program consists of the \texttt{main} function and two threads.
The \texttt{main} function first initializes the I2C driver and configures the bus speed.
The two drivers we used for this work are BCM \cite{bcmdriver} and Linux i2c-dev \cite{linuxi2c} (version 4.4).
In the rest of this paper we refer to these drivers simply as \textit{BCM} and \textit{Linux}.
%this is i2c-tools version 3.1.2-3
Next, INA219's configuration registers are programmed to adjust gain and resolution.
The programmed gain value determines the maximum measurable current, and resolution refers to the number of bits per sample.
The \texttt{main} function then initiates two threads: the \texttt{sampler} thread polls the conversion ready bit, and when set, it reads the bus voltage and shunt voltage values.
In addition, the current system time (in nanoseconds) is read using the \texttt{clock\_gettime()} system call.
As this system call is used to add timing information to each sample, it is important to ensure its delay is negligible.
To this end, we used a high sampling rate (500MHz) logic analyzer and observed that the pin toggling delay using BCM library is 40ns.
Then, we toggled the pin right before and after the system call. 
This study showed that the delay of this system call is around 445ns with negligible variations; therefore, this provides a reliable mechanism to append a time stamp to each sample collected, which is later used for energy calculation.

The \texttt{sampler} thread also computes energy consumption through calling the \texttt{compute\_energy()} function, which implements the Riemann integral approach.
% by computing the area of trapezoids. 
In addition to energy calculation, raw data are written to a file.
To this end, we introduced the \texttt{sample\_writer} thread to read data off a buffer and write to a file.
We employed two different buffering mechanisms to implement \textit{batched} and \textit{continuous} file write operations: 
When using the \textit{two-buffer} mechanism, the \texttt{sampler} thread and \texttt{sample\_writer} thread use two different buffers. 
When a buffer is full, the \texttt{sampler} thread unblocks the \texttt{sample\_writer} to flush that buffer. 
Therefore, the entries are batched and then written to the file. 
Meanwhile, the \texttt{sampler} thread uses the other buffer.
The other approach utilizes a single \textit{circular buffer}.
In this approach, as soon as a new entry is written to the buffer, the \texttt{sampler} thread signals the \texttt{sample\_writer} thread to write that entry to the file.
In this case, a mutex locks the buffer to avoid concurrent accesses.
For both cases the buffers are implemented as fixed-size arrays allocated on stack memory to avoid the overhead of memory re-allocation.
This also results in a higher utilization of cache memory due to the contiguous placement of array entries in random access memory.

The software supports raw data collection and energy measurement: (i) for a given time duration, (ii) after a certain number of samples were collected, and (iii) by receiving an external trigger to indicate the start and stop of measurement.
In the trigger mode, falling and rising edge interrupts trigger the starting and stopping of power measurement.
Using this feature, one can annotate an IoT device's code to start and stop the measurement at particular locations, thereby providing fine-grained energy measurement of various operations such as encryption and transmission.

\begin{algorithm}[t]
	\footnotesize
	\SetInd{0.9em}{0.7em}

    \SetKwFunction{FMain}{main}
    \SetKwProg{Fn}{function}{}{}
    \Fn{\FMain{}}
    {
        setup the I$^{2}$C driver (BCM/Linux) and bus speed\;
        configure the INA219 gain and resolution\;
        create thread \texttt{sampler}\;
        create thread \texttt{sample\_writer}\;
        \KwRet\;
    }

    \SetKwFunction{FMain}{sampler}
    \SetKwProg{Fn}{thread}{}{}
    \Fn{\FMain{}}
    {
        \While{sampling is enabled}
        {
        
        \While {conversion ready == 0} {check the bit\;}
            
            read bus voltage and shunt voltage values\;
            
            \texttt{compute\_energy}(\texttt{prev\_smp}, \texttt{new\_smp});
        
            \uIf {use two buffers}
            {
                \uIf {\texttt{use\_buffer1} == true }
                {
                    add entry to \texttt{buffer1}\;
                    \uIf{\texttt{buffer1} is full} 
                    {
                        signal the \texttt{sample\_writer} thread\;
                        \texttt{use\_buffer1} == false\;
                    }
                }
                \Else 
                {
                    add entry to the \texttt{buffer2}\;
                    \uIf{\texttt{buffer2} is full} 
                    {
                        signal the \texttt{sample\_writer} thread\;
                        \texttt{use\_buffer1} == true\;
                    }
                }
            }
            \uElseIf{use a circular buffer}
            {
                lock the buffer\;
                add entry to the buffer\;
                unlock the buffer\;
                signal the \texttt{sample\_writer} thread\;
            }
        }
        
        % \KwRet\;
    }

    \SetKwFunction{FMain}{sample\_writer}
    \SetKwProg{Fn}{thread}{}{}
    \Fn{\FMain{}}
    {
        \uIf {use two buffers}
        {
            \uIf {\texttt{use\_buffer1} == true }
            {
                flush \texttt{buffer1} to the file;
            }
            \Else
            {
                flush \texttt{buffer2} to the file;
            }
        }
        \uElseIf{use a circular buffer}
t        {
            lock the buffer\;
            write entry to the file\;
            unlock the buffer\;
        }
    }
    
    \SetKwFunction{FMain}{compute\_energy}
    \SetKwProg{Fn}{function}{}{}
    \Fn{\FMain{\texttt{prev\_smp}, \texttt{new\_smp}}}
    {    
        \texttt{prv\_power} = (\texttt{prv\_smp.voltage}) $\times$ (\texttt{prv\_smp.current})\;
        \texttt{new\_power} = (\texttt{new\_smp.voltage}) $\times$ (\texttt{new\_smp.current})\;
        \texttt{time\_diff} = (\texttt{new\_smp.time}) $\times$ (\texttt{prv\_smp.time})\;

        \texttt{new\_energy} = \texttt{new\_power} $\times$ \texttt{time\_diff} \;
        \texttt{tri} = (\texttt{new\_power} $-$ \texttt{prv\_power}) $\times$ \texttt{time\_diff} $/$ 2  \;
        \texttt{new\_energy} -= \texttt{tri}\;

        \texttt{energy} += \texttt{new\_energy}\;

    }
    
	\caption{EMPIOT's Software}
	\label{alg:empiot_sw}	
\end{algorithm}

\subsection{Design Parameters}
In this section we study the effect of various design parameters on performance in terms of sampling rate, sampling offset, accuracy, and energy consumption of the base board.

\subsubsection{Sampling Rate}
\label{eff_sampl_rate}
\begin{figure}[!t]
\centering
\includegraphics[width=1\linewidth]{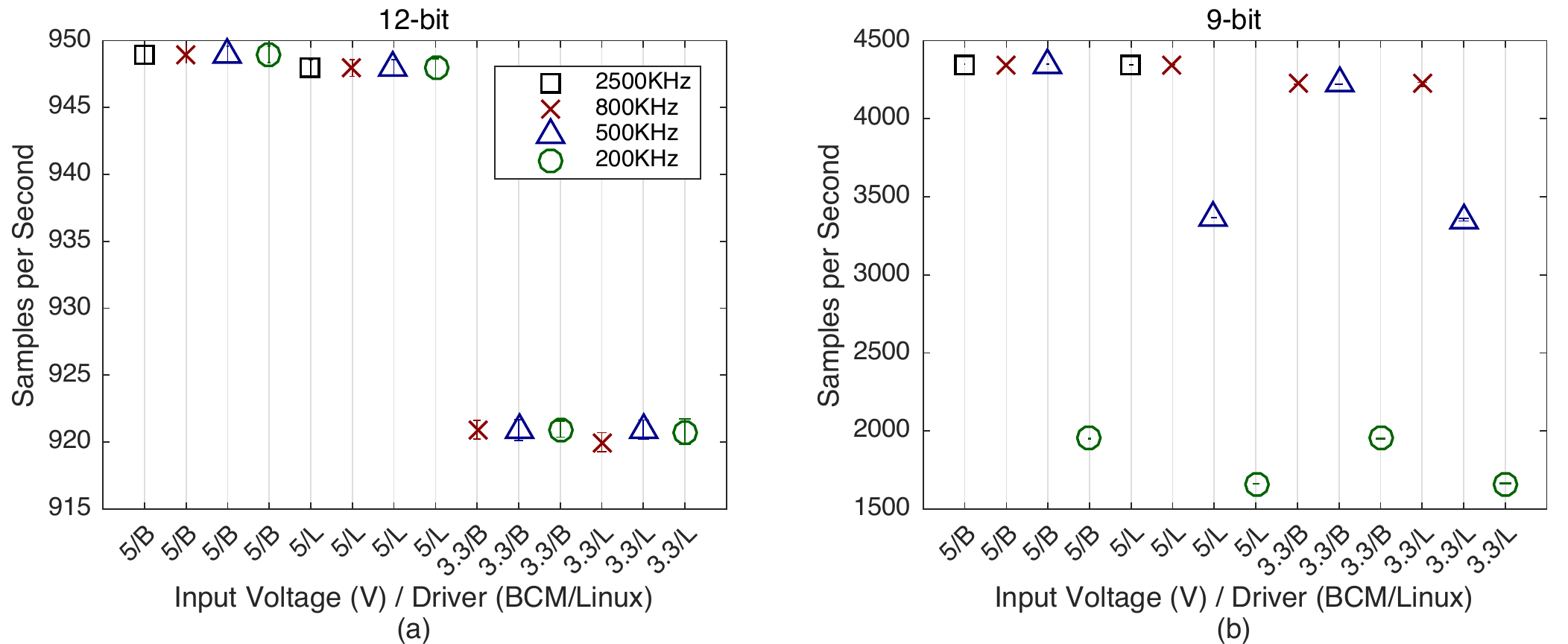}
\caption{Sampling rate for (a) 12-bit resolution and (b) 9-bit resolution. Error bars show 95\% confidence interval.}
\label{fig:12_bit_sampling}
\end{figure}

Figure \ref{fig:12_bit_sampling} shows sampling rate versus bus speed, driver, and input voltage, where "input voltage" refers to the power source of EMPIOT's shield.
The main observations are as follows: First, the sampling rate is lower than the conversion rate supported by the chip.
For example, for 12-bit conversion, the sampling  rate is lower than the 1.8KHz ADC conversion rate we reported in Section \ref{hw}.
Second, a lower voltage results in a lower sampling rate.
Third, reducing the bus speed to as low as 200KHz results in reducing the sampling rate.
Fourth, the Linux driver affects sampling rate for 9-bit resolution.
We study these observations in more details as follows.

\begin{figure}[!t]
\centering
\includegraphics[width=0.67\linewidth]{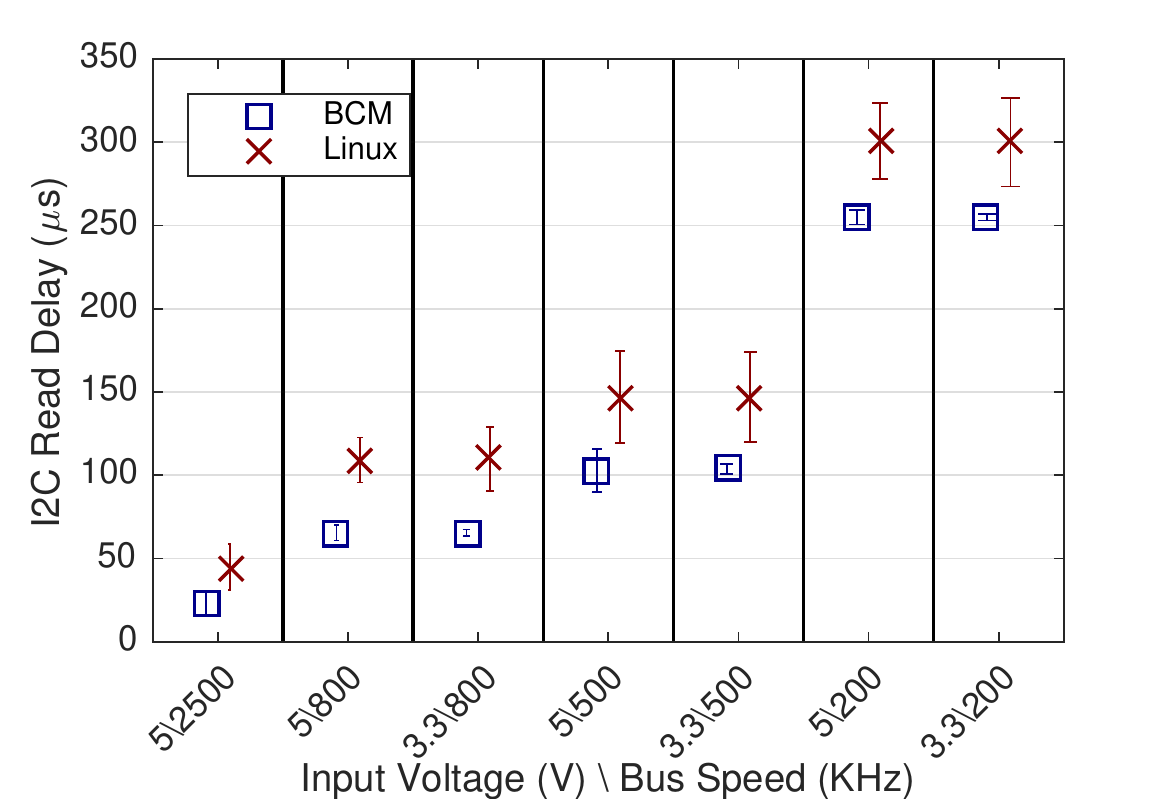}
\caption{I2C read delay measured as the time interval between issuing a read command and reception of requested bytes. Error bars show 95\% confidence interval.}
\label{fig:bcm_vs_linux_i2c_read}
\end{figure}

In order to evaluate software overhead, we measured the time spent by the \texttt{sampler} thread between collecting a sample and the next polling of conversion ready bit.
Using a high-speed logic analyzer, our results show that the processing overhead of the \texttt{sampler} thread is 0.46$\mu$s, which is negligible.
%In fact, we observed that the interval between setting the conversion ready bit is about 1.02 ms and 228 $\mu$s for 12-bit and 9-bit sampling, respectively.
%Although the datasheet reports ADC conversion time, it does not include the actual delay of setting and clearing the conversion ready bit. 
%
\begin{figure}[!t]
\centering
\includegraphics[width=0.75\linewidth]{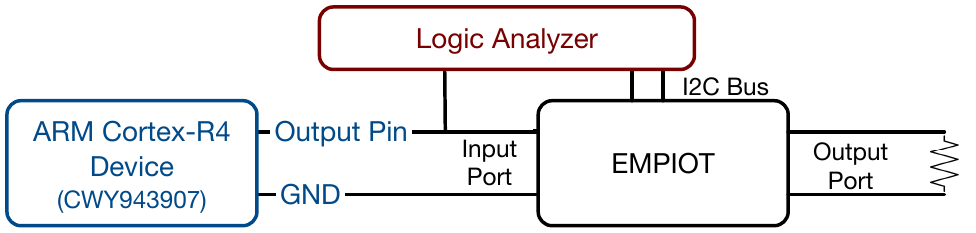}
\caption{In order to measure I2C read delay, the logic analyzer captures I2C communications.
To measure sampling delay offset, the logic analyzer captures both "Output Pin" status and I2C communications. }
\label{fig:delay_measurement_circuit}
\end{figure}
To measure I2C read delay, we used a simple program that continuously reads two bytes from the shield board.
After capturing I2C traffic using a logic analyzer, we compute the interval between sending I2C addresses.
Figure \ref{fig:bcm_vs_linux_i2c_read} shows the results\footnote{Please note that we did not report the results for bus speed 2500KHz and input voltage 3.3V because we observed a very unreliable I2C communication in this condition. We believe that INA219 cannot keep up with this high clock rate when the voltage is 3.3V.}.
This figure reveals the effect of bus speed and driver on read delay.
In particular, the BCM driver achieves a faster and more stable I2C performance compared to the Linux driver.
In fact, for a given baud rate, the Linux I2C read is at least 20$\mu$s slower than BCM.
% For example, for the bus speed 2.5MHz, the interval between sending an I2C address and reading the bytes is about 1.8$\mu$s and 12$\mu$s when using the BCM and Linux driver, respectively.
In order to justify the lower performance of Linux driver compared to BCM, we used \texttt{strace} \cite{strace} to capture the system calls made by the software.
Our evaluations show that when using BCM, the number of system calls is always fixed (exactly 181) and does not depend on the sampling rate.
We also observed that these system calls are only made during the initialization of the BCM driver, after which the software directly communicates with the driver.
In contrast, when using the Linux driver, all the I2C communication requests pass through the kernel; therefore, the number of system calls depends on the sampling rate.
%These over justified the higher performance of using the BCM driver.
%Therefore, when using the Linux driver, the higher delay is due to the (i) longer delay of starting I2C communication when a system call is made, (ii) the longer interval between sending two consecutive I2C commands, (iii) the longer delay of returning from a system call.
Figure \ref{fig:bcm_vs_linux_i2c_read} also shows that the I2C delay of Linux driver has higher variations compared to BCM.
The average range of variations for Linux I2C read delay is 22$\mu$s, and for BCM this value is 4$\mu$s.
These results indicate that the Linux driver does not achieve a stable communication rate with mission-critical and high rate sensors, such as those used in medical and industrial applications.
The effect is also obvious when using 9-bit sampling (cf. Figure \ref{fig:12_bit_sampling}(b)).
For example, when using 500KHz bus speed, using the Linux driver reduces the sampling rate to 3360, compared to the 4350 samples collected per second when the BCM driver is in use.
From the software point of view, the overhead of Linux driver affects the number of times the conversion ready bit is polled per sample collection round.
Table \ref{bcm_vs_linux} reports the results.
Therefore, the Linux driver falls behind the sample conversion rate when the sampling rate is high.

\begin{figure}[!t]
\centering
\includegraphics[width=0.7\linewidth]{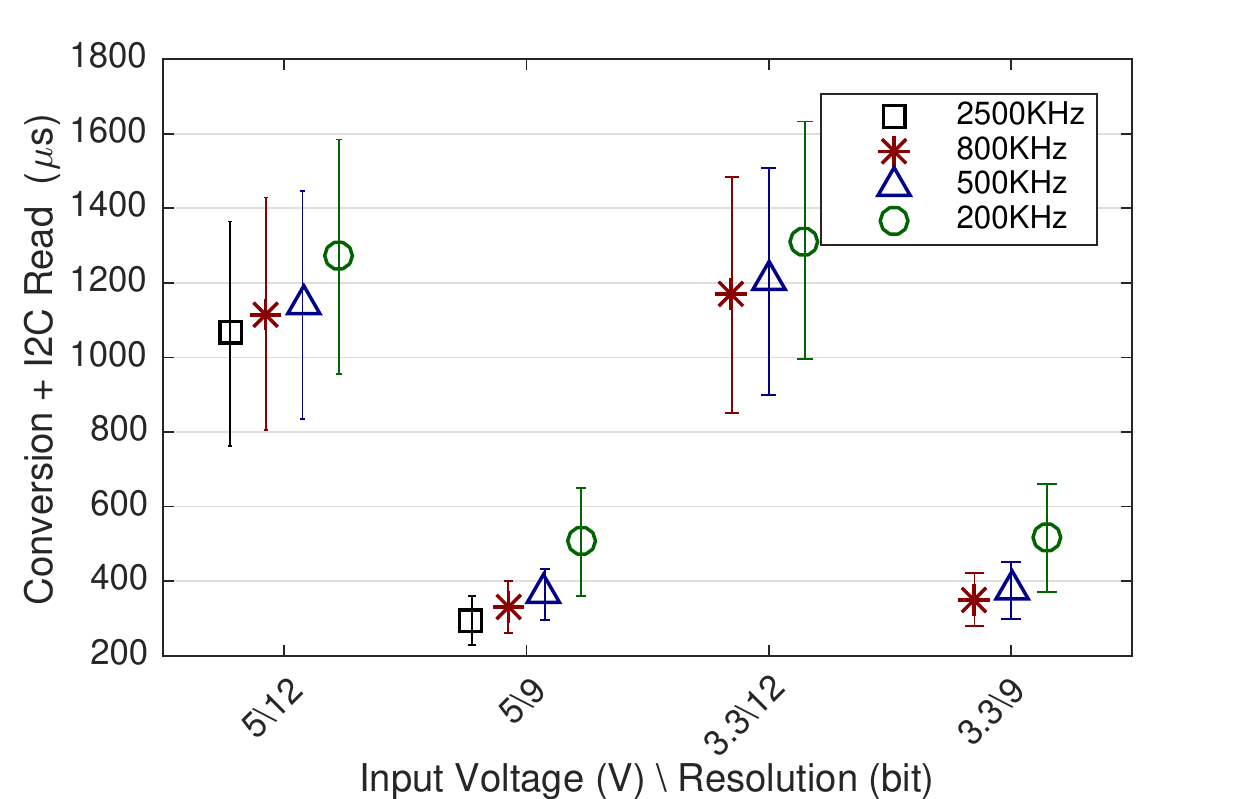}
\caption{Sampling offset, which reflects the interval between a change in input current and reading the corresponding value. We used BCM driver for this experiment. Error bars show 95\% confidence interval.}
\label{fig:conversion_time_delay}
\end{figure}

We next analyzed the time interval between a change in input and reading the corresponding value, which is referred to as \textit{sampling offset}.
In order to identify the causes of sampling offset, we used the experiment shown in Figure \ref{fig:delay_measurement_circuit}.
%In other words, we determine the time offset of each sample, which would be beneficial when the user needs to know the exact time a change in input has happened.
% In order to decompose I2C read delay from hardware sampling delay, we also measured the delay between a change in power voltage and the time the change is reported (i.e., when the conversion ready bit is set).
To introduce quick and predictable changes in power, we have used a pin toggle using an ARM-Cortex R4 board (CWY9443907 \cite{CYW43907}) which sets a pin from high to low (i.e., 3.3V to 0V) in 50ns.
For this experiment, the "Output Pin" in Figure \ref{fig:delay_measurement_circuit} is initially high (3.3V).
At particular intervals, the pin is set to low (0V) and remains in this state for 2ms.
We compute the sampling delay offset by measuring the interval between setting a pin to low and collecting the corresponding sample.
To this end, a logic analyzer logs the status of "Output Pin" as well as I2C communications.
Figure \ref{fig:conversion_time_delay} shows the measured values.
Since the I2C read delay is independent of the input voltage (as Figure \ref{fig:bcm_vs_linux_i2c_read} shows), the results of Figure \ref{fig:conversion_time_delay} indicate a longer conversion time when using a lower voltage value, i.e., 3.3V.
For example, increasing conversion time by 64$\mu$s decreases the number of samples collected per second by 47.
As the power measurement chip uses a delta-sigma ADC, we believe that the lower voltage value slows down the operation of the decimator and averaging circuitry.

\begin{table}[]
\footnotesize
\centering
\caption{Polling rate of conversion ready bit using BCM and Linux drivers for 12 and 9-bit resolution.}
\label{bcm_vs_linux}
\setlength{\tabcolsep}{0.3em} % for the horizontal padding
\begin{tabular}{cc|c|c|c|c|}
\hhline{~~|-|-|-|-|}
                                              &       & \multicolumn{4}{c|}{Polling Rate}   \\   \hhline{~~|-|-|-|-|}
                                              &       & {\cellcolor[gray]{\ShadingColor}}2500KHz & {\cellcolor[gray]{\ShadingColor}}800KHz  & {\cellcolor[gray]{\ShadingColor}}500KHz  & {\cellcolor[gray]{\ShadingColor}}200KHz \\ \hline \hline
\multicolumn{1}{|c|}{\multirow{2}{*}{12-bit}} &  {\cellcolor[gray]{\ShadingColor}}BCM   & 45  & 15    & 9 & 3 \\ \hhline{|~|-|-|-|-|-|}
\multicolumn{1}{|c|}{}                        & {\cellcolor[gray]{\ShadingColor}}Linux & 23  & 9      & 6    & 2      \\ \hline \hline
\multicolumn{1}{|c|}{\multirow{2}{*}{9-bit}}  & {\cellcolor[gray]{\ShadingColor}}BCM   & 9  & 2     &
1   & 1  \\ \hhline{|~|-|-|-|-|-|}
\multicolumn{1}{|c|}{}                        & {\cellcolor[gray]{\ShadingColor}}Linux & 4  & 1    & 1      & 1 
\\ \hline
\end{tabular}
\end{table}

\subsubsection{Measurement Accuracy}

In this section we analyze the effect of input voltage variations on accuracy.
We consider the following voltage sources to run the INA219 of EMPIOT's shield board:
\begin{itemize}
    \item External (5VExt): A 5V external power \cite{U8001}.%
    \item 5V from RPi (5VRPi): The 5V pin of RPi header. 
    \item 5V from RPi with a load (5VRPi w/Load): The 5V pin of RPi header. 
    In addition, the RPi's USB port is connected to a Cypress CYW943907 IoT device. The idle energy consumption of this device is about 100mA, but it periodically wakes up and sends ping packets that result in up to 400mA current consumption.
    \item 3.3V from RPi (3.3VRPi): The 3.3V pin of RPi header.
    %Note that for these experiments we did not use 3.3V w/Load as its variations is lower than that of 5 V.
\end{itemize}
We have used an industrial-grade DMM \cite{kth_dmm_7510} with a sampling rate of 500Ksps to measure the variations of these power sources.
The 95\% variations of these sources are as follows: 5VExt: 2mV, 5VRPi: 103mV, 5VRPi w/Load: 300mV, 3.3VRPi: 24mV.
Caused by operating system processes and EMPIOT software, we observe that RPi as a power source exhibits higher variations compared to 5VExt.
Nevertheless, in order to simplify the design and reduce its cost, we are interested in confirming if RPi can be used as the source of power for EMPIOT's shield board.
%
%
% \begin{table}[]
% \centering
% \caption{The stability of various voltage sources}
% \label{voltage_var}
% \begin{tabular}{c|c|c|c|c|}
% \hhline{~|-|-|-|-|}

%                                       & {\cellcolor[gray]{\ShadingColor}}{Max}    & {\cellcolor[gray]{\ShadingColor}}{Min}    & {\cellcolor[gray]{\ShadingColor}}{Mean}   & {\cellcolor[gray]{\ShadingColor}}{2$\times\sigma$ }   \\ \hline
%  \multicolumn{1}{|c|}{{\cellcolor[gray]{\ShadingColor}}{5V-External}}     & 5.0032 & 5.0024 & 5.0028 & 0.00022 \\ \hline 
% \multicolumn{1}{|c|}{{\cellcolor[gray]{\ShadingColor}}{5V-RPi}}          & 4.9641 & 4.8665 & 4.9347 & 0.021   \\ \hline
% \multicolumn{1}{|c|}{{\cellcolor[gray]{\ShadingColor}}{5V-RPi w/Load}}   & 4.9346 & 4.8631 & 4.9246 & 0.027   \\ \hline
% \multicolumn{1}{|c|}{{\cellcolor[gray]{\ShadingColor}}{3.3V-RPi}}        & 3.2857 & 3.2852 & 3.2854 & 0.00014 \\ \hline
% \multicolumn{1}{|c|}{{\cellcolor[gray]{\ShadingColor}}{3.3V-RPi w/Load}} & 3.2857 & 3.2851 & 3.2854 & 0.00019 \\ \hline
% \end{tabular}
% \end{table}

\begin{figure}[!t]
\centering
\includegraphics[width=1\linewidth]{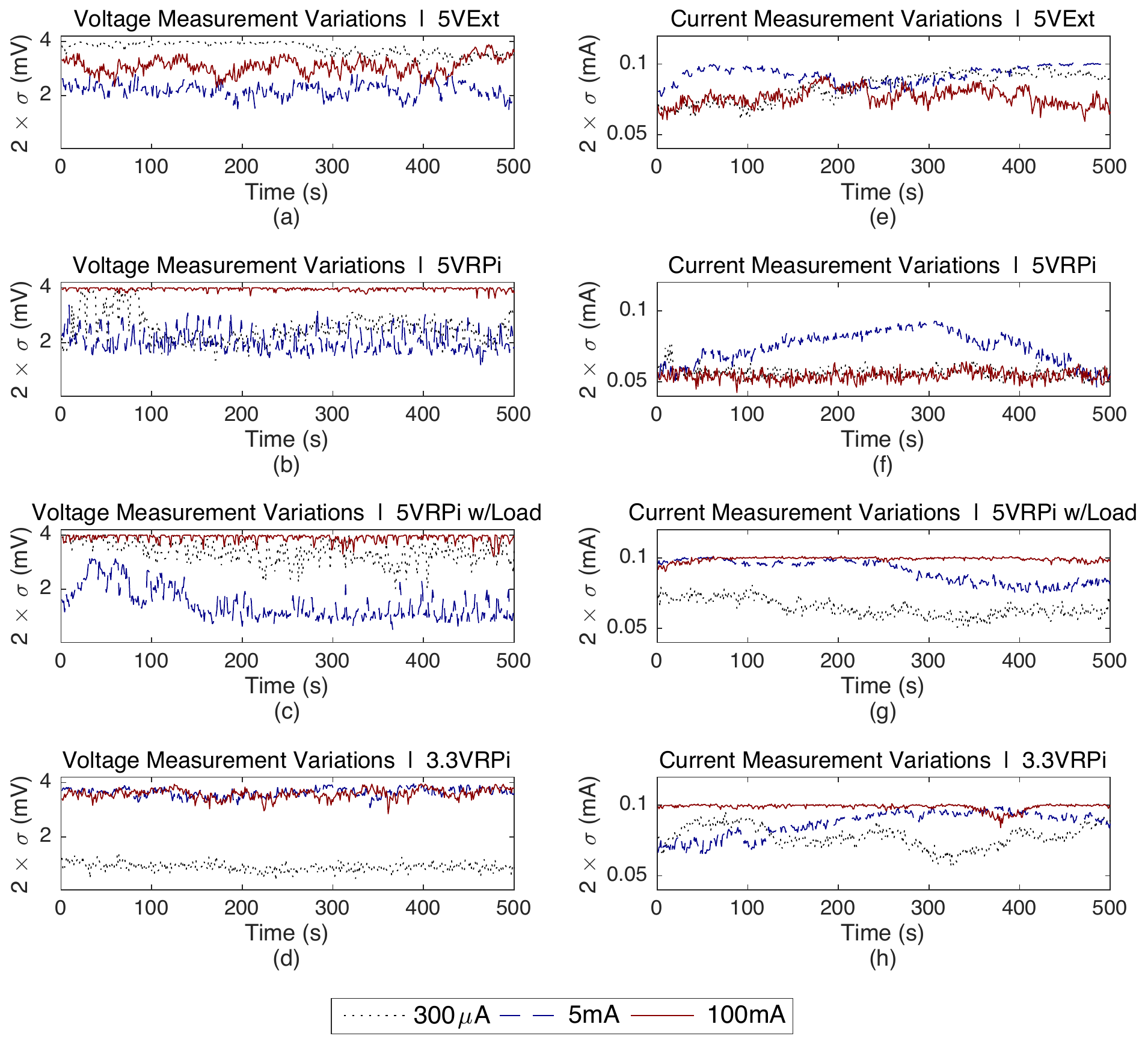}
\caption{The measurement variations of voltage (sub-figure (a) through (d)) and current (sub-figure (e) through (h)) over time for three different loads and various input sources. Sampling resolution is 12 bit. The y-axis is the 95\% range of variations.}
\label{fig:voltage_var_accuracy}
\end{figure}

In order to measure accuracy across a wide range, we used three different fixed loads: 200$\mu$A, 5mA and 100mA. 
These loads are generated by connecting the shield's output to three different resistors.
Figure \ref{fig:voltage_var_accuracy} presents the effect of voltage variation on the accuracy of bus voltage and current measurement.
These results show the variability of measurements caused by factors such as electromagnetic interference and white noise.
However, the important observation is that, irrespective to the source of power and load value, the measurement error of EMPIOT is always less than 4mV and 0.1mA for bus voltage and current, respectively. 
The reported error ranges comply with the values we previously mentioned for INA219 (cf. Section \ref{hw}).

%
%BUS SPEED IS 1MHZ

{\color{black!50!black}
\subsubsection{Overhead Analysis}
\label{overhead}
In this section we investigate the energy consumption of the RPi running EMPIOT's software.
The two main causes of power consumption are file write and polling the shield board.
In these experiments we measure the energy consumption of EMPIOT boards using either RPi3 (Raspberry Pi 3) or RPiZW (Raspberry Pi Zero with WiFi)\footnote{The RPi3 used in this paper is based on a 900MHz BCM2837, and the RPiZW is based on a BCM2835 SoC.}.
In addition to energy, we have logged the time of writing each sample to the file through toggling a pin and logging pin activation times.
This technique avoids introducing extra software overhead to log file write instances.
To remove the variations caused by Ethernet/wireless communication, we used UART to communicate with the RPi board whose energy is being measured.
For example, not only are power variations reduced when using UART instead of Ethernet, but current consumption is reduced by about 30mA as well.
In our implementation, each sample entry is 16 bytes: 8 bytes to store timestamp (i.e., the \texttt{timespec} structure), 4 bytes for bus voltage, and 4 bytes for current.

% Figure \ref{fig:buffering_effect_rpi3} and \ref{fig:buffering_effect_rpizw} show the results for RPi3 and RPiZW, respectively.
Figure \ref{fig:buffering_effect_rpi3} shows the power consumption trace for RPi3.
Vertical red bars indicate file write activities.
As it can be observed, using the circular buffer mechanism results in continuously writing to the flash memory, which in turn increases the energy consumption.
For example, for 9-bit resolution and BCM driver, the base power is increased by about 0.05w when using the circular buffer mechanism (compare sub-figure (a) and (c)).
We have extracted a similar set of traces for RPiZW, and we observed that this increase is about 0.008W for this board.

Using the traces collected for RPi3 and RPiZW, Figure \ref{fig:buffering_effect_summary} summarizes the effect of sampling resolution, driver, and buffering mechanisms on energy consumption.
In order to prepare this figure, although the power consumption of the circular buffer mechanism is readily available, we need to compute the energy consumption of the two-buffer mechanism based on the energy consumption of polling and file write, as follows.
The \texttt{sample\_writer} thread is activated when a buffer is full.
We refer to the time required to fill a buffer as $t_{b}$.
After $t_{b}$ seconds of buffering samples, the \texttt{sample\_writer} thread is activated to write the samples to a file.
It should be noted that, during file write, the device is also buffering samples.
We refer to the file write duration as $t_{wb}$.
Based on these values, we can compute the average energy consumption per second as 
\begin{equation}
\begin{split}
    E = E_{b} + E_{wb} = p_{b} \times \frac{t_{b} - t_{wb}}{t_{b}}  +  p_{wb} \times \frac{t_{wb}}{t_{b}}  \\
\end{split}
\end{equation}
where $p_{b}$ is the power consumption of buffering operation, and $p_{wb}$ is the power consumption of writing and buffering simultaneously.
The two variables $t_{b}$ and $t_{wb}$ are computed as follows.
The duration of filling a buffer of size $L_{b}$ samples is $t_b = L_{b}/R_{s}$, where $R_{s}$ is the sampling rate.
The \texttt{sample\_writer} thread requires $t_{w} = L_{b} \times L_{s} / W $ seconds to transfer $L_{b}$ samples to a file, where $W$ is the file write speed in bits per second and $L_{s}$ is the length of each sample in bits.
Therefore, the energy consumption of the two-buffer mechanism is
\begin{equation}
\begin{split}
    E = \frac{p_{b} \times W + L_{s}\times R_{s} \times (p_{wb} - p_{b}) }  {W}  \\
\end{split}
\end{equation}
As Figure \ref{fig:buffering_effect_summary} shows, the preferred buffering mechanisms for RPi3 and RPiZW are two-buffer and circular buffer, respectively.
Since RPi3 has a multi-core processor, it schedules the \texttt{sample\_writer} thread on a different core than that of the \texttt{sampler} thread.
Although scheduling the \texttt{sample\_writer} thread on a separate core increases processor power consumption, the speed of file write process is increased.
On the other hand, RPiZW has a single core, so, the \texttt{sample\_writer} thread requires more time to write the buffered data to the file.
The longer $t_{wb}$ duration of RPiZW increases the impact of $p_{wb}$.
However, in the circular buffering mode the processor writes small amounts of data between polling instances, so the energy consumption of circular buffer is less than the two-buffer mechanism.

% Since the \texttt{sample\_writer}'s processor core utilization is around 98\%, this increase requires the utilization of another processor core, thereby preventing core sleep.

Figure \ref{fig:buffering_effect_summary} also reflects the higher energy consumption of the Linux I2C driver.
For example, on RPi3, when using two-buffer and 12-bit resolution, using the Linux driver increases the base power to 1.38W, as compared to 1.26W achieved with the BCM driver.
As we discussed earlier, I2C communication through the Linux driver results in a significantly higher number of system calls, which increases processing load.

% In this paper we used a RPi3 board as the EMPIOT's based board.
% However, when Ethernet port and high processing power are not required, cheaper and more power-efficient boards such as Raspberry Pi Zero with Wi-Fi (RPiZW) can be employed.
% Our studies show that for a RPiZW with an active Wi-Fi connection, using the BCM driver and batched write operations results in 670mW power consumption.
% Using the circular buffer and Linux driver increases power consumption to 718mW and 685mW, respectively.

% Although we have used a small buffer size (20000) for these experiments to reveal the effect of batched write operations, a much larger buffer size results in a higher energy saving.
% Even if only 30\% of the 1 GB RAM of RPi3 or 25\% of the 512MB of RPiZW is used, then the number of entries kept in random access memory before each file write will be 19,660,000 and 8,388,608, respectively.
% This shows that batching can be easily used to minimize energy consumption.

%
%

% -------------------------Bus speed is 2MHz.
\begin{figure}[!t]
\centering
\includegraphics[width=0.9\linewidth]{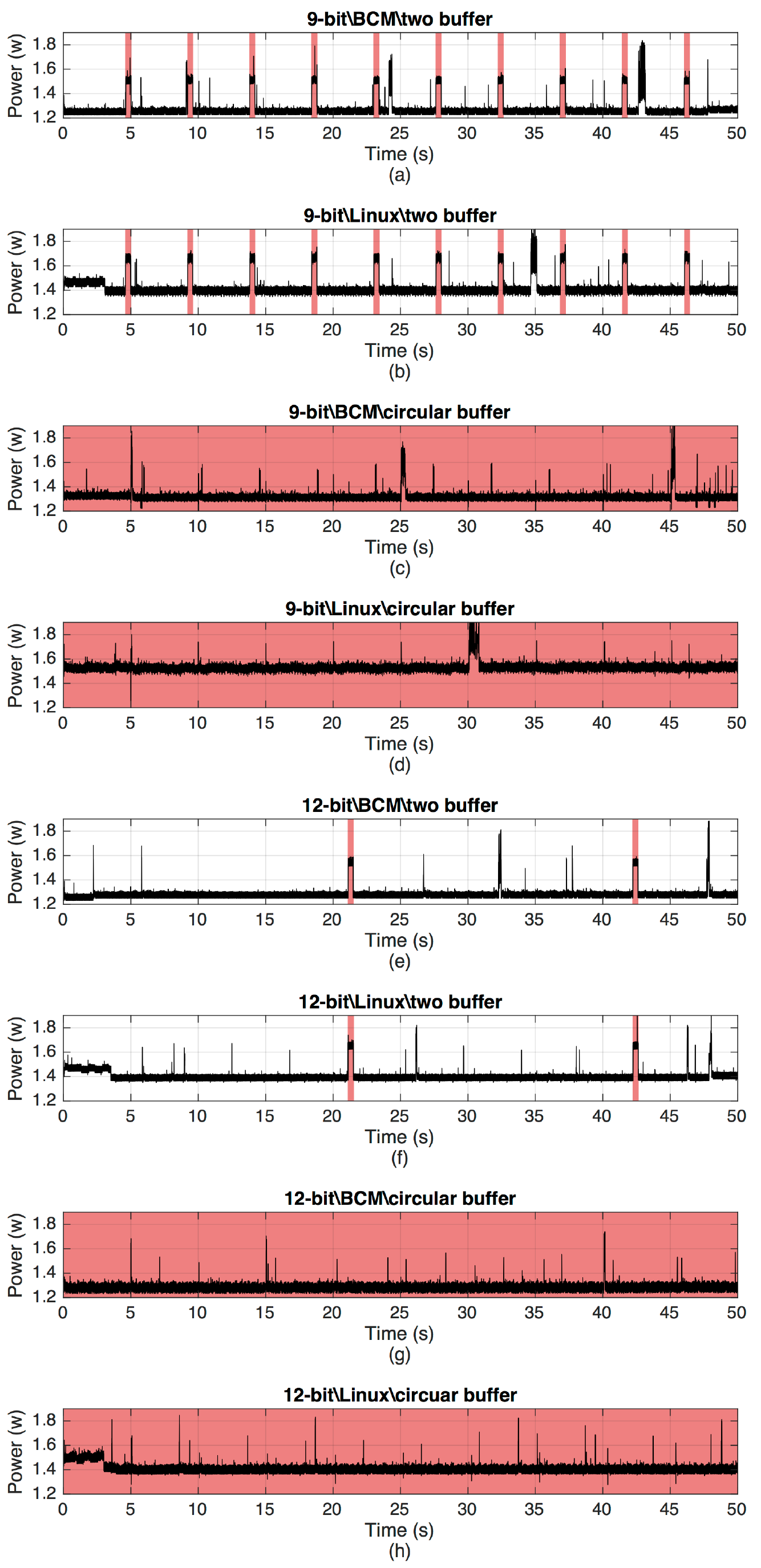}
\caption{The power trace of RPi3 versus sampling resolution, driver, and buffering mechanisms.}
\label{fig:buffering_effect_rpi3}
\end{figure}

\begin{figure}[!t]
\centering
\includegraphics[width=1\linewidth]{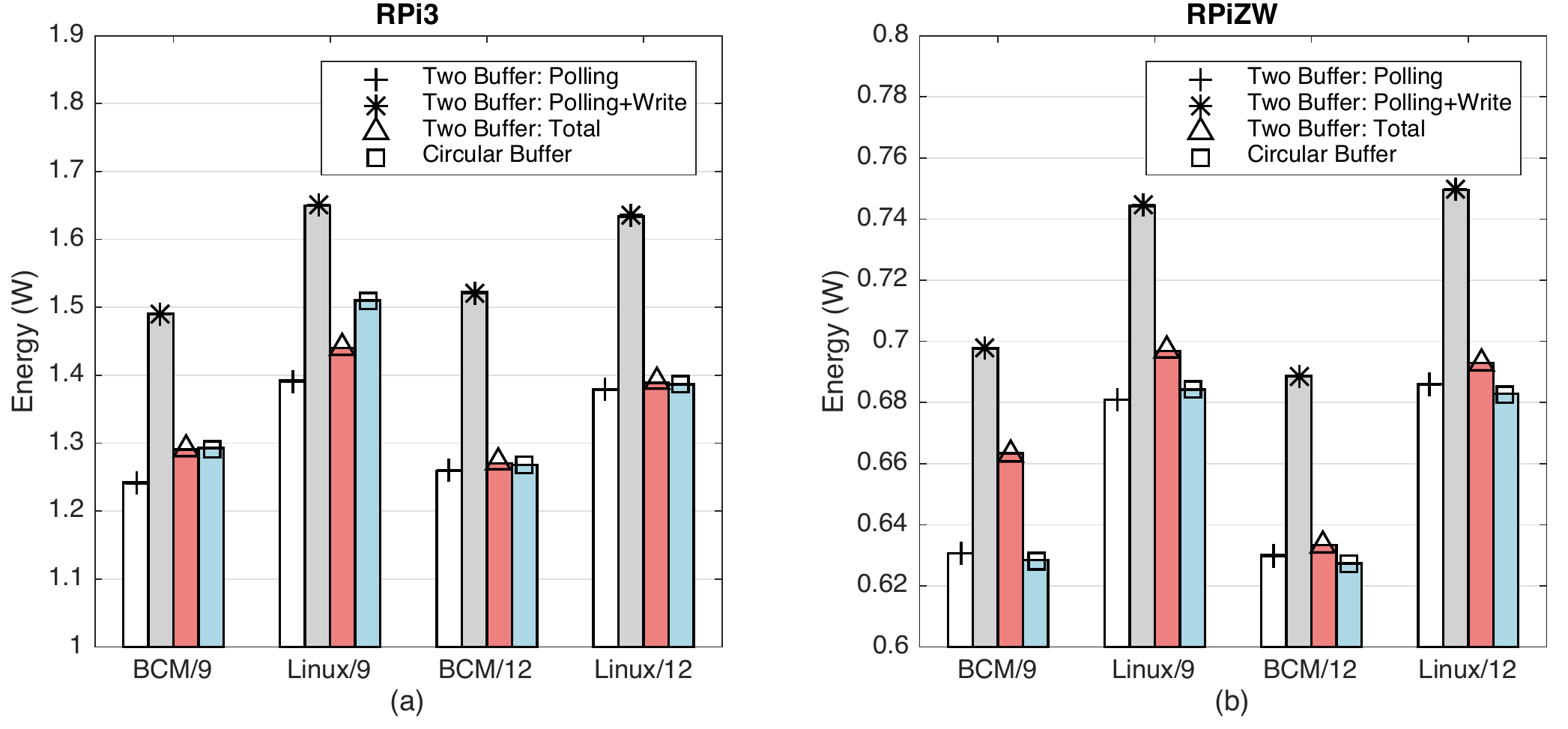}
\caption{The effect of sampling resolution, driver, and buffering mechanisms, on the energy consumption of RPi3 and RPiZW running EMPIOT software per second.}
\label{fig:buffering_effect_summary}
\end{figure}

}

{\color{black!50!black}
\subsection{Overcoming the Two Limitations of EMPIOT}
\label{limitations}
As mentioned earlier, the current measurement resolution of the proposed platform is 100$\mu$A.
However, since IoT devices support multiple low-power modes, current draws less than 100$\mu$A cannot be measured.
% For example, EMPIOT cannot capture transitions between 1$\mu$A, 10$\mu$A and 100$\mu$A low-power states.
Our solution to this problem is to use \textit{a hybrid energy calculation model} that computes the energy consumption of low-power modes similar to analytical energy estimation models.
Specifically, we require the IoT device to inform EMPIOT right before and after each transition into a low-power mode.
In this case, energy is computed as follows,
\begin{equation}
\begin{split}
\sum_{\forall p_{ps}[i] \in \mathbf{p_{ps}}}\;\; \sum_{\forall (t_{s}, t_{e}) \in p_{ps}}^{} (t_{e} - t_{s}) \times p_{ps}[i] +
\sum_{\forall S_{j} \in \mathbf{S}} \Delta_{j} p_{j}
\end{split}
\end{equation}
where $\mathbf{p_{ps}}$ is the set of all power saving modes that have a power consumption less than $100\mu$A, $p_{ps}[i]$ is power saving mode $i$, $t_{s}$ is the start of power saving mode, $t_{e}$ is the end of power saving mode, $\mathbf{S}$ is the set of samples collected during the normal operation (i.e., no power saving mode), $S_{j}$ is a sample $j$ collected during normal operation, and $\Delta_{j}$ and $p_{j}$ refer to the duration and power of sample $S_{j}$, respectively. 
Please note that during the power saving modes, power samples (i.e., $S_{j}$) are not used for energy measurement.
This technique of capturing low power consumption is accurate and easy to implement due to two reasons:
First, using a constant value as the power consumed in a low-power mode is reasonable because these modes refer to sleep states during which power consumption is stable.
Second, the number of low-power states is usually limited to two or three modes.
For example, CC2650 \cite{CC2650} offers two sleep modes: \textit{standby} and \textit{shutdown}.
During the standby mode, real-time clock (RTC) is running and the contents of RAM and CPU are retained.
The device wake ups from the shutdown mode through a trigger.
The energy consumed by these two modes are 1$\mu$A and 100nA, respectively. 
For those devices that the power consumption of their sleep modes is not known \textit{a priori}, a DMM could be used to extract the respective values, which are then hardcoded into EMPIOT.
We will evaluate the effectiveness of the hybrid energy estimation technique in Section \ref{perf_eval}.

% In order to show the effectiveness of this approach, for the experiments conducted in Section \ref{perf_eval}, we used the standby mode of CC2650.
% Through generating GPIO interrupts, we inform the EMPIOT board when the device enters and leaves the satndby mode.
% Our experimental results show that this mechanism results in less than 4\% energy measurement error, as Figure \ref{fig:baeline_comapre}(a) indicates.

Another limitation of EMPIOT is its warm-up time.
Our studies show that the first 3 to 5 samples collected after initialization are not reliable.
Therefore, when using 12-bit resolution, the first 5ms of an operation cannot be measured.
This limits the minimum duration of energy measurement.
This problem, however, can be simply addressed in software: instead of actually turning on and off the shield board per measurement, we can ignore the samples collected when energy measurement is inactive.
}

\section{Calibration}
\label{calib}
{\color{black!50!black}
Factors such as the resistance of the shunt resistor path, inaccuracy of shunt resistor, and ADC non-linearity cause differences between the EMPIOT's measurements and a ground truth.
Thereby, calibration is an essential part of the design.
To this end, the purpose of current calibration is to find function $f_{A}(i_{e})$ so that
\begin{equation}
i_{a} = f_{A}(i_{e}) \;\;\;\;\; 0 \leq i_{a} \leq i_{max}
\end{equation}
where $i_{a}$ is the actual current draw, $i_{e}$ is the current reported by EMPIOT when the actual current is $i_{a}$, and $i_{max}$ is the maximum current draw of the IoT device.
Similarly, the purpose of current calibration is to find function $f_{V}(v_{e})$ so that
\begin{equation}
v_{a} = f_{V}(v_{e}) \;\;\;\;\; 0 \leq v_{a} \leq v_{max}
\end{equation}
where $v_{a}$ is the actual voltage, $v_{e}$ is the voltage reported by EMPIOT when the actual voltage is $v_{a}$, and $v_{max}$ is the maximum voltage supported by IoT device.
}

Accurate estimation of  $f_{A}(i_{e})$ and  $f_{V}(v_{e})$ requires a load that can generate current and voltage values across the supported measurement range of EMPIOT.
Although current variations can be generated by using an IoT device, this method does not result in an accurate calibration due to the following reasons:
First, the duration of a change in current draw may not be long enough to match the samples collected. 
In other words, due to the fast variations of current as well as the difference in the sampling offset of the DMM and EMPIOT, correlation of the samples collected by the two devices is very challenging.
For example, when the current draw changes suddenly, some values may not be captured by EMPIOT due to its lower sampling rate compared to a DMM.
Furthermore, DMM and EMPIOT might report the variations with different time offsets. 
Therefore, subtracting the pairwise values does not reflect a realistic measurement error.
If the error of EMPIOT versus DMM is in fact a linear function, then the aforementioned calibration errors may prevent us from finding a linear fit or cause a linear function that its slope is higher or lower than the real value.
In addition, the IoT board may not cover the supported current range of EMPIOT, which results in calibration gaps.
In fact, since turning on each component (such as RF transceiver) results in a jump in energy consumption, it is almost impossible to generate a linear increase.
Another solution is to use a potentiometer as the load.
However, similar to using an IoT device, we cannot predict the transition and duration of drawing a particular current value, so the measurement offsets affect calibration error.
Furthermore, potentiometers usually support low current values, and the calibration for currents higher than 100mA requires an expensive potentiometer.
Due to these limitations, the existing works perform calibration either: (i) manually by using fixed resistor values \cite{Haratcherev2008,Duttac2008,Trathnigg2008}, or (ii) by using expensive equipment \cite{Lim2013a,Potsch2017,Naderiparizi2016}.

{\color{black!50!black}
To address the aforementioned challenges and simplify the calibration process, we have designed a low-cost, accurate, and \textit{programmable calibration tool} which provides dynamic voltage and current ranges.
This calibration tool is in fact a programmable load where its resistance and timing characteristics are controllable through a software (written in Python) running on a RPi.  
This software controls reconfiguration frequency and records output settling time between two consecutive configurations. 
Specifically, if reconfiguration frequency is $t$, the software records the output settling times at $n \times t + t/2$, where $n \in Z^{+}$ corresponding to all of the supported output values.
Recording output settling instances enables us to correlate the measurement values of DMM and EMPIOT when the load is stable. 
Furthermore, this feature prevents the need for accurate time synchronization of DMM and EMPIOT.

The calibration tool uses an 8-bit digital potentiometer, AD5200~\cite{AD5200}, which has a maximum resistance of 10k$\Omega$. 
The maximum current output of this digital potentiometer can be expressed as $I_{potmax}={V_{in}}/{R_{w}}$, where $I_{potmax}$ is the maximum current supported by the digital potentiometer when it is programmed to the minimum value, $R_{w}$ is the constant wiper resistance when the digital potentiometer is programmed to 0, and $V_{in}$ is the voltage input to the calibration tool.
Since the resistance range of the digital potentiometer is limited, we have added a number of resistors in parallel in order to extend the supported current range.
Therefore, based on Kirchoff's law, the total maximum current of calibration tool is expressed as follows,
\begin{equation}
  \label{eqn:current_sum}
    \left. I_{max} = I_{potmax} + \sum_{j=1}^n I_j \;\;\;\; \forall n \in Z^+ \right.
\end{equation}
where $I_{max}$ is the maximum supported current of the calibration tool, $\sum_{j=1}^n I_j$ is the sum of the currents that $n$ resistors can support, and $ I_j$ is the current that a resistor $R_j$ carries.
In our design, when $V_{in}=5V$, the maximum current output of calibration tool is 1A.
To dynamically adjust the resistance of the calibration tool, resistors $R_{j}$ are attached to a switch network that is implemented by four digital ADG1612~\cite{ADG1612} switches.
Therefore, although the digital potentiometer cannot handle currents higher than 20mA, our design supports high current ranges by switching on and off the resistor paths.
Specifically, a new resistor path is enabled whenever a 20mA or a 100mA increase in current is required, and the digital potentiometer is used to fine-tune current between range 0 to 20mA.
By programming the resistance value, the calibration tool can generate various current and voltage values. 
The calibration tool's software programs AD5200 and ADG1612 through SPI and GPIO interfaces.

The minimum current resolution supported by the calibration tool, denoted as $I_{res}$, depends on two parameters of the digital potentiometer:
maximum resistance $R_{max}$, and the number of programmable bits.
The equation that determines the digitally-programmed output resistance is expressed as follows,
\begin{equation}
  \label{eqn:digipot resistor}
   \left. R(x) = \frac{x}{2^n}\times R_{max}+R_w  \;\;\;\;\;\;  0\leq x \leq 2^n \right.
\end{equation}
where $x$ is the value programmed into the digital potentiometer, $n$ is the number of bits supported by digital potentiometer, $R_w$ is the constant wiper resistance, and $R_{max}$ is the digital potentiometer's maximum resistance.
Therefore, the minimum resolution of current output is expressed as follows,
\begin{equation}
  \label{eqn:minimum res}
   \left.  I_{res} = \frac{(R(x)-R(x-1))}{R(x)\times R(x-1)} \times V_{in} \;\;\;\;\;\; 0 < x\leq 2^n \right.
\end{equation}
Based on the components used and Equation \ref{eqn:minimum res}, the minimum current output is $0.476mA$ and the minimum current resolution is $1.82\mu$A.
In addition, minimum voltage output is $0.06mV$ and minimum voltage resolution is $0.01\mu$V. 
The wide range and high resolution of the calibration tool enable us to calibrate both current and voltage within the operating range of various IoT devices.
It is worth mentioning that the entire calibration solution costs about \$100, including manufacturing costs. 
This is less than 1\% of the cost compared to current commercial solutions.
% By using a high-accuracy digital potentiometer and well-calculated resistor combinations, the output range of ProCal spans between $0.476mA$ to 1A for current and $0.06mV$ to 5.5V for voltage.
}

The calibration tool generates line interrupts to trigger the start and stop of measurements by EMPIOT and DMM.
As mentioned earlier, EMPIOT can be started and stopped through line interrupts.
To trigger the DMM, we have used a high-accuracy and programmable device \cite{kth_dmm_7510}, which exposes several programmable GPIO pins.
We have developed a Python script using SCPI (Standard Commands for Programmable Instruments) to configure the sampling rate and enable the DMM to start and stop sampling based on the line interrupts received.
This script communicates with the DMM through a TCP/IP connection and transfers the sampled data from the DMM buffer to a PC when a measurement completes.

EMPIOT's shield includes a jumper to enable current flow from input to output (cf. Figure \ref{fig:empiot_hw}).
For calibration purposes we have removed this jumper and connected the pins to a DMM; thereby, EMPIOT and DMM measure current simultaneously.
Instead of pairwise comparison of the traces collected by EMPIOT and DMM, we use the timing data logged by the programmable load.
As the timing data reflects load stability instances, we can safely compare the two closest entries of the traces collected by DMM and EMPIOT.
%For example, for each timing value $t_{i}$, we find the entries in DMM and EMPIOT, $t^{minDMM}_{i}$ and $t^{minEMPIOT}_{i}$, where their timing distance from $t_{i}$ is minimal.

One of the goals of this paper is to show if an off-the-shelf INA219 breakout board can be used instead of EMPIOT's shield with the software and configuration parameters proposed in this work to achieve a high level of accuracy.
To this end, in addition to reporting calibration data for EMPIOT's shield, we have used an INA219 breakout board \cite{ina219adafruit} to study the effect of hardware design on calibration.
This breakout board has been installed on a bread board and communicates with a RPi running EMPIOT's software.

\begin{figure}[!t]
\centering
\includegraphics[width=1\linewidth]{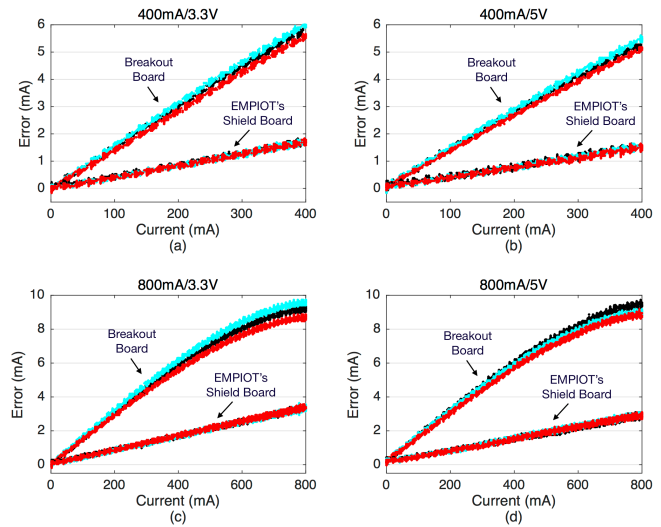}
\caption{Current measurement error ($i_{a} - i_{e}$) versus ground truth. The value above each figure shows the maximum supported current (left side) and input voltage (right side).
}
\label{fig:meas_error_calib}
\end{figure}

Figure \ref{fig:meas_error_calib} shows the measurement errors of three EMPIOT boards and three breakout boards conducted in a normal indoor temperature 25$^{\circ}$C.
Please note that the value above each figure refers to: (i) left value: the maximum supported current configured through the programmable gain amplifier (PGA), and (ii) right value: the input voltage.
As it can be observed, EMPIOT's shield presents lower error compared to the breakout boards.
Since INA219 measures current through a shunt resistor, the distance and impedance of the circuit path between the resistor and chip highly affect measurement accuracy.
Therefore, the higher accuracy of EMPIOT's shield is due to the thicker and shorter path used.
In terms of input voltage, using 5V instead of 3.3V slightly reduces error. 
In addition, Figure \ref{fig:meas_error_calib} shows that the error of EMPIOT's shield increases linearly versus current.
However, the breakout boards' error shows a quadratic behavior for currents beyond 300mA.
For both these cases, instead of using a calibration table, we simply find the best fitted curves.
Table \ref{cal_values} reports the calibration values for these boards.

%also shows that higher voltage results in lower error.

%
%
%
\begin{table*}[ht]
\centering
	\def\arraystretch{1.2}
\caption{Current Calibration Functions}
\label{cal_values}
{\color{black!50!black}
\begin{tabular}{|l|c|c|c|c|}
\hline
\multirow{4}{*}{\rotatebox[origin=c]{90}{\centering EMPIOT}} & 400mA/3.3V &  $i_{e} = f_{A}^{-1}(i_{a}) = 0.9957 \; i_{a}$ & RMSE < 0.00005 & R-Square = 1  \\ \cline{2-5} 
& 400mA/5V  & $i_{e} = f_{A}^{-1}(i_{a}) = 0.9963 \; i_{a}$ & RMSE < 0.00005 & R-Square = 1 \\ \cline{2-5} 
& 800mA/3.3V  & $i_{e} = f_{A}^{-1}(i_{a}) = 0.9949 \; i_{a}$ & RMSE < 0.0024 & R-Square > 0.9998 \\ \cline{2-5} 
& 800mA/5V   & $i_{e} = f_{A}^{-1}(i_{a}) = 0.9956 \; i_{a}$ & RMSE < 0.0016 & R-Square = 1 \\ \hline\hline

\multirow{4}{*}{\rotatebox[origin=c]{90}{\centering \makecell{Breakout\\ Board}}} & 400mA/3.3V &  $i_{e} = f_{A}^{-1}(i_{a}) = 0.9853 \; i_{a}$  &  RMSE < 0.00006 & R-Square = 1 \\ \cline{2-5} 
& 400mA/5V  & $i_{e} = f_{A}^{-1}(i_{a}) = 0.9869 \; i_{a}$ & RMSE < 0.00006 & R-Square = 1\ \\ \cline{2-5} 
& 800mA/3.3V  & $i_{e} = f_{A}^{-1}(i_{a}) = 0.0079 \; i_{a}^{2} + 0.9816 i_{a}$ & RMSE < 0.0028 & R-Square > 0.9998 \\ \cline{2-5} 
& 800mA/5V   & $i_{e} = f_{A}^{-1}(i_{a}) = 0.0074 \; i_{a}^{2} + 0.982 i_{a}$ & RMSE < 0.0061 & R-Square > 0.9994 \\ \hline

\end{tabular}
}
\end{table*}
{\color{black!50!black}
% voltage calibration
In addition to current, the voltage measured may not reflect the actual bus voltage.
We used the calibration tool to generate a variable voltage in the range 2V to 5.5V.
Our results show that the error of voltage measurement is a fixed offset.
In fact, EMPIOT's voltage measurements versus DMM results in a linear function $v_{a} = f_{V}(v_{e}) = v_{e} + 0.027$.
% The R-square and RMSE of this fitting are 1 and 0.0016, respectively.
For the breakout board the calibration function is $v_{a} = f_{V}(v_{e}) = v_{e} + 0.097$.
%However, compared to the EMPIOT board, any changes to the wiring results in changes in the calibration value.
}

\section{Performance Evaluation}
\label{perf_eval}

In this section we study the effectiveness of design parameters and calibration on accuracy when EMPIOT's shield and the breakout board are used.
Based on the results reported in Section \ref{platfDesign}, the following setting is used for both EMPIOT's shield and the breakout board: (i) BCM driver is used, (ii) bus speed is 2500KHz, (iii) INA219's voltage is 5V, and (iv) file writes are batched.
Our ground truth is the energy measured by two high accuracy DMMs \cite{kth_dmm_7510} that record current and voltage with 500Ksps sampling rate and 18-bit resolution.
A Python script programs the DMMs in trigger modes.
Therefore, when an IoT device begins its operation, EMPIOT as well as the two DMMs start energy measurement.
The three main operations performed by the IoT devices are \textit{sleep}, \textit{software encryption} and \textit{transmission}. 
Please note that before each transmission the transceiver listens to the medium due to employing carrier-sense multiple access (CSMA) \cite{dezfouli2015modeling}.
The encryption operation is used to generate a high processing load.
Depending on the workload used, each node transitions between the available states every 500ms.
We introduce four types of loads:
\begin{itemize}
    \item Workload 1: All the three operations are included,
    \item Workload 2: Includes send and sleep operations,
    \item Workload 3: Includes encryption and sleep operations,
    \item Workload 4: Includes encryption and send operations.    
\end{itemize}

We have used five different IoT boards with various energy characteristics.
{\color{black!50!black}
The first device is a TI SensorTag CC2650 \cite{CC2650}.
The CC2650 SoC includes an ARM Cortex-M3 processor and supports IEEE 802.15.4 standard.
Since the energy consumption of this device in sleep mode is 1$\mu$A, we used this board to measure the accuracy of EMPIOT for profiling the energy of very low-power devices.
Specifically, we are interested in seeing the effectiveness of the hybrid energy measurement technique proposed in Section \ref{limitations} in terms of accuracy.
We used TI-RTOS \cite{TI_RTOS} as the operating system running on this device.
During the transmission state, the device sends 30-byte 802.15.4 packets as fast as possible.
The small packet size introduces quick variations in power consumption.
}

In order to generate very fast and high temporal variations in power, we used four 802.11-based IoT devices: Avnet BCM4343W \cite{BCM4343W}, Cypress CYW43907 \cite{CYW43907,CYW43907AEVAL}, RPiZW, and RPi3.
The Avnet board includes an ARM Cortex-M4 processor and supports 802.11a/g/n.
When using the power save mode, the minimum energy consumption of the board is around 10mA, processing consumes around 40mA, and packet transmission results in spikes up to 350mA.
The Cypress board includes an ARM Cortex-R4 processor and supports 802.11a/g/n.
As the board includes other components like an Ethernet chip, the sleep power of the board is around 96mA. 
The processing power is around 140mA, and packet transmissions increase power consumption up to 400mA.
We have used Free-RTOS and WICED Studio \cite{WICED} for software development on the Avnet and Cypress devices.
Furthermore, to generate higher variations in power, the software developed for these devices enables the power save mode (PS-Poll) mechanism of 802.11. 
At certain intervals the radio wakes up and ping packets are transmitted as fast as possible.
Since ping packets are small, this behavior results in fast and short spikes in power consumption.

For RPiZW, the base and processing power are about 130mA and 180mA, respectively, and 802.11 transmissions result in spikes as high as 300mA.
For the regular RPi3, the sleep and processing currents are about 280mA and 330mA, respectively, and 802.11 transmissions increase current consumption up to 500mA. 
A program (written in c language) controls the transition between the three operations provided.
Please note that for RPiZW and RPi3 we refer to the process inactivity time as sleep time.

\begin{figure}[!t]
\centering
\includegraphics[width=1\linewidth]{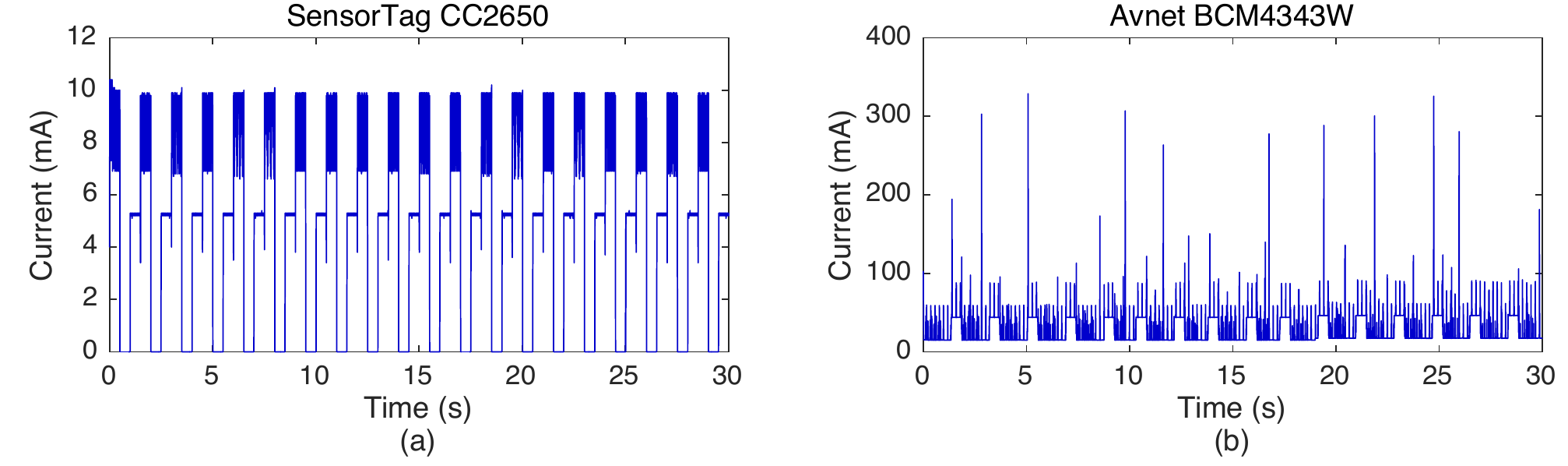}
\caption{The energy trace of SensorTag CC2650 (using 802.15.4) and Avnet BCM4343W (using 802.11) when transitioning between sleep, encryption and transmission. 
This figure shows the significantly higher variations of current consumption caused by 802.11 compared with 802.15.4.
}
\label{fig:sample_workload}
\end{figure}

Figure \ref{fig:sample_workload} shows the energy consumption of CC2650 and BCM4343W.
Although CC2650 shows a clear transition between the three states, BCM4343W presents spikes across the trace.
It should be noted that in contrast with 802.15.4, 802.11 communication requires association with an access point.
Therefore, when 802.11 power save mode is enabled, the device wakes up every 100ms to receive the beacon packets generated by the access point.

%

%
% \begin{figure}[!t]
% \centering
% \includegraphics[width=1\linewidth]{IoT-Energy-Measurement/figures/ECDF_current.pdf}
% \caption{The empirical cumulative distribution function of current samples.
% }
% \label{fig:ecdf_current}
% \end{figure}
%
%
%

\begin{figure}[!t]
\centering
\includegraphics[width=1\linewidth]{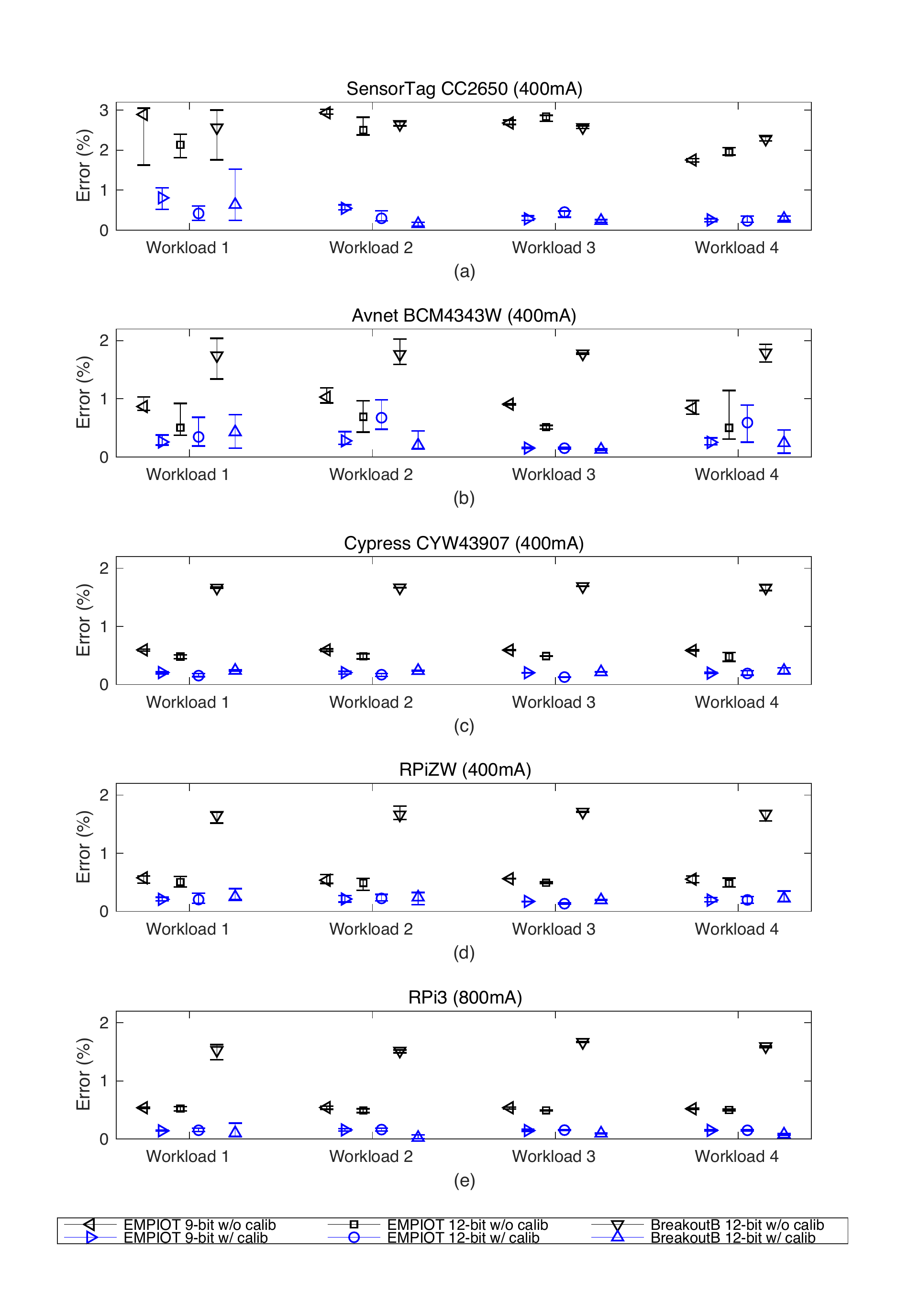}
\caption{{\color{black!50!black}The energy measurement error of EMPIOT versus the ground truth. Error bars show median, lower quartile and higher quartile. Each marker is the median of ten experiments each 30 seconds long.}
}
\label{fig:baeline_comapre}
\end{figure}

Figure \ref{fig:baeline_comapre} presents the power measurement error when using EMPIOT's shield and the breakout board.
Error is computed as $\frac{ \left | E_{x} - E_{a} \right |  }{E_{a}}\times 100$, where $E_{a}$ is the actual energy consumption measured by DMM, and $E_{x}$ refers to the energy measured by either EMPIOT's shield or the breakout board.
Each marker is the median of 10 experiments, where an experiment is 30 seconds long\footnote{The 30-second experiment duration is due to the limitation of the DMM used in terms of the number of samples stored in its memory.}.
These results indicate that EMPIOT is in fact an accurate power measurement platform, even if an off-the-shelf breakout board is used instead of the shield.
Comparing sub-figure (a) through (e) shows that energy measurement error is higher when measuring very small currents.
{\color{black!50!black}
Specifically, while the measurement error for 802.11-based boards is less than 2.5\%, the error is less than 3.5\% for SensorTag.
Considering Workload 1 on SensorTag, we observed that more than 40\% of variations in current are less than 1mA. % as \label{fig:ecdf_current}(a) shows.
If the variation is caused by transition to the sleep mode, the hybrid power measurement model is used and the CC2650 device informs EMPIOT about its transition to and from the sleep mode by generating interrupts.
However, since the current resolution of INA219 is 100$\mu$A, for example, EMPIOT cannot detect current variations between 3.25mA and 3.27mA.
Therefore, as the power consumption of SensorTag is significantly lower than that of other boards, small errors in current measurement result in a more considerable effect on total measurement error.
Nevertheless, these results confirm the effectiveness of the hybrid power measurement technique.
}

Figure \ref{fig:baeline_comapre} also reveals the effect of calibration on accuracy.
Since the measurement error of the breakout board is higher than that of EMPIOT's shield, calibration has a higher effect on accuracy.
In addition, this is particularly important for 802.11-based boards because, as Figure \ref{fig:meas_error_calib} shows, the error of the breakout board is higher and non-linear for currents higher than 300mA.

Although our studies in Section \ref{eff_sampl_rate} showed that EMPIOT supports around 1000 and 4000 samples per second for 12 and 9-bit resolution, respectively, it should be noted that the actual sampling rate of the delta-sigma ADC is about 500KHz, and the decimator averages the analog samples collected per sampling interval.
For example, when 12-bit sampling is used for 802.11-based devices, the rapid power variations of these platforms are captured and averaged per millisecond, according to the sampling rate  we reported in Section \ref{eff_sampl_rate}.
Consequently, we observe that EMPIOT achieves high accuracy even in the presence of very fast temporal power variations.

Figure \ref{fig:baeline_comapre} also shows that the measurement error of 9-bit resolution is slightly higher than that of 12-bit resolution.
This is particularly obvious for the SensorTag measurements. 
As we mentioned earlier, compared with the 802.11 devices, SensorTag generates smaller variations in power, therefore a higher resolution is required to capture the changes.
% It is worth noting that the measurement error of EMPIOT is due to two reasons: (i) lower sampling resolution, and (ii) lower sampling rate, compared to the DMM.
Compared with 12-bit, using 9-bit resolution enhances the sampling rate (cf. Figure \ref{fig:12_bit_sampling}), however, the minimum detectable variation in current is increased to 780$\mu$A.
For 802.11-based devices with a high data transmission rate, the higher sampling rate enables us to capture shorter variations in power.
Therefore, the increase in sampling rate compensates the error introduced due to lower resolution.

\subsection{Importance of Voltage Measurement}
\begin{figure}[!t]
\centering
\includegraphics[width=0.65\linewidth]{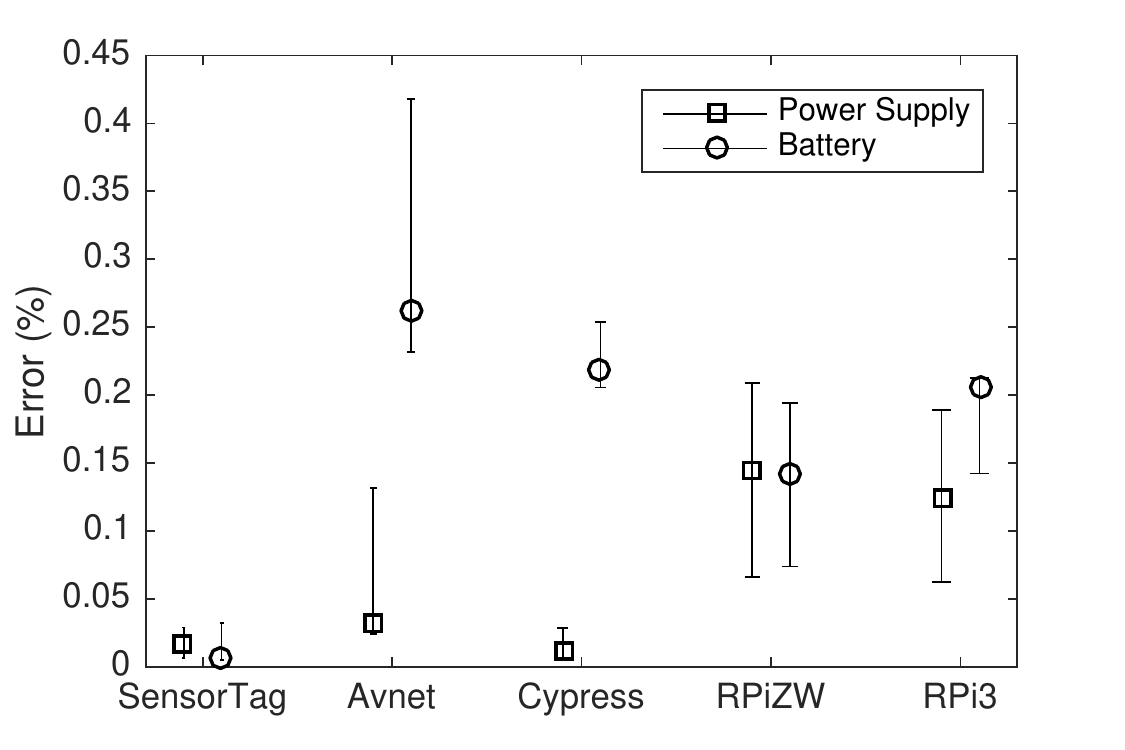}
\caption{The effect of ignoring voltage on the accuracy of energy measurement. Error bars show median, lower quartile and higher quartile.
}
\label{fig:voltage_effect}
\end{figure}

Most of the existing energy measurement platforms ignore the effect of voltage variations on energy measurement (e.g., \cite{Haratcherev2008,Andersen2009,Zhou2013a,Trathnigg2008}).
To verify the implication of this assumption on accuracy, we have used Workload 1 and computed energy when:
(i) the average value of voltage measurements is used, and (ii) the voltage samples are used.
Error is computed as the difference between these cases.
Figure \ref{fig:voltage_effect} shows the results when a power supply and a battery are used as the sources of power for various types of IoT devices.
These results indicate that neglecting voltage would result in up to a 0.45\% increase in energy measurement error.
The impact of voltage on energy measurement depends on various factors including: the stability of power source, the number and intensity of sudden increases in current draw, and the electronic characteristics of the IoT device such as voltage stabilization. 
For example, when a battery is used, sudden variations of current result in a higher measurement error when voltage is ignored.
These results also show that the Cypress and Avnet devices cause significant variations in input voltage, compared with RPi3 and RPiZW.
Our studies show that these variations are caused by the operations of 802.11 transceiver, thereby highlighting the importance of including voltage for 802.11-based IoT devices.
On the other hand, the energy consumption of SensorTag and its 802.15.4 transceiver does not cause any significant variation in voltage.

\section{Related Work}
\label{related_work}
In this section we provide an overview of the existing energy measurement solutions in two domains: empirical measurement, and analytical modeling.

% \subsection{Power Measurement}
% The most popular approach of current measurement is to use a low-value shunt resistor and measure the voltage drop across the resistor.
% Give the low value of shunt resistor, the shunt voltage would be very low.
% However, since ADCs usually work in the range 0 to 5V, the shunt voltage should be amplified.
% A \textit{current sense amplifier} (CSA) (a.k.a., current shunt monitors) can be used.
% The most important parameter of a CSA is its gain, which indicates the ratio of output voltage to input voltage. 
% The gain may be either fixed, or adjustable depending on the value of the resistor attached to the chip.
% The other important parameter is the offset voltage at the DC input of the CSA.
% In particular, since low current values result in very low shunt volatile, the total error will be affected by the amplifier's offset voltage.

\subsection{Empirical Measurement}
\label{related_emp}
Based on their main shortcoming, we categorize the platforms proposed for empirical energy measurement as follows.

\subsubsection{Costly and Bulky}
An oscilloscope is used in \cite{feeney2001investigating} to analyze the energy consumption of 802.11 transceivers.
%Although using oscilloscopes' sampling rate is high (Gsps), they usually provide milliamp-level sensitivity, which is not enough to reflect minor changes in power.
Using a current probe is the approach employed in \cite{Milenkovic2005} for power monitoring.
The current probe relies on the magnetic field generated by current draw; therefore, it cannot be used to detect small currents or variations (milliamps level).
The authors in \cite{Kaup2014} added a shunt resistor to a USB cable and measured both current and bus voltage using USB1608-FSPlus \cite{MCC}, a 16-bit ADC.
The USB1608-FSPlus is controlled by a PC and the sampling rate is 1Ksps.
%They also presented the mathematical modeling of power measurement error as a function of ADC and shunt resistor error.
%This works studies the energy consumption of the board versus CPU utilization, Ethernet data rate, and WiFi bandwidth.
All of these approaches are expensive and their size prevents integration with the IoT devices of a testbed.

\subsubsection{Complex Circuit}
%--------------------------------
% [-] requiring building a circuit
Similar to ICs such as BQ2019 \cite{ti_bat_mon}, the SPOT \cite{Jiang2007} platform uses voltage to frequency conversion and relies on the resources of the device under test to operate. 
In addition, the oscillator and the converter are sources of noise and error and may interfere with the device under test.
%****  The dynamic range provided is 45000:1 with 1$\mu$A resolution.
More importantly, attaching SPOT to an IoT device is not plug-and-play.
SPOT's measurement error is less than 15\% and its resolution is 1$\mu$A; however, it assumes that the typical operating range of the IoT device is 5$\mu$A to 50mA. 
%Also, instead of streaming the measured values, SPOT measures energy consumption over a time period.
%--------------------------------
% [-] requiring building a circuit
LEAP2 \cite{Stathopoulos2008} is a FPGA-based platform that enables individual monitoring of various components such as processor and memory. %with sampling rates up to 25KHz.
% In this architecture, a high bandwidth memory bus enables the sharing of collected energy data with the host processor. 
%Prior work [13, 1, 9] has demonstrated that a more capable CPU (such as the PXA255 or PXA270), while incurring significantly higher peak power dissipation than an MCU (e.g., MSP430) is in fact more energy efficient while un- der load, due to its architectural advantages—higher clock speed, L1 caches, hardware MMU, multi-stage pipeline etc. Therefore, our goal for this set of experiments was to verify whether a CPU running at a higher speed is still more energy efficient than the same CPU running at a lower speed.
%--------------------------------
% [-] requiring building a circuit
iCount \cite{Duttac2008} relies on the linear relationship between current and switching frequency of boost switching regulators; counting the switching cycles reflects the current drawn during an interval.
A shortcoming of iCount is that it utilizes the resources of device under test and increases its energy consumption.
%In addition, due to the reading overhead of the counter, the maximum sampling rate is 66 kHz. 
%However, the switching frequency adjusts to a new rate within about 125 $\mu$s.
Another limitation of this approach is that boost converters are not always available on IoT boards. 
For example, both CYW43907 \cite{CYW43907} and CC2650 \cite{CC2650} (used in Section \ref{perf_eval}) utilize internal voltage regulators.
% Unfortunately, the authors did not show the accuracy of iCount in a real measurement.
%--------------------------------

%2009
% [-] requiring building a circuit
%Quote: As sensor node current consumption ranges over 5 decades (1 μA – 100 mA), this implies that a (linear) ADC must have a voltage resolution of 18 bits or more,
Energy Bucket \cite{Andersen2009} counts the number of charges and discharges of a buffer capacitor.
%The charge and discharge voltages of the capacitor are precisely controlled.
Although Energy Bucket can measure currents in the range 1$\mu$A to 100mA, the major shortcoming is the dependency of sampling rate on capacitor value.
For example, when the current drawn is very small, the inter-sampling interval would be long.
In addition, the platform assumes that bus voltage is fixed, and the paper does not include accuracy analysis. 
%The paper does not provide any energy measurement in a real and complicated workload.
%Also, the accuracy and speed of the platform have not been cross-validated.
%--------------------------------
% [-] requiring building a circuit
% [-] does not measure voltage
%2013
%they mention high resolution and high accuracy are easy to achieve
%By referring to the studies on \cite{Jiang2007}, they mention that the spikes caused by switching between different states (sleep, awake) of TelosB \cite{TelosB} node are between 200 to 400 $\mu$s.
% Quote: Nemo must achieve a dynamic range of 100,000:1 (from 2 uA to 200 mA)
% ******** IMPORTANT: Therefore, a sampling rate 5KHz or faster is required to capture these spikes.
% Quote: When Nemo is connected to a PC to upload measurements, a sampling rate of 100 KHz can be achieved by disabling the compressing algorithm.
Nemo \cite{Zhou2013a} uses a shunt resistor switch composed of a series of resistors.
Each resistor is enabled or disabled based on current intensity.
The voltage across the resistor switch is amplified using a differential op-amp and then digitized by a 12-bit ADC.
% Quote: Two popular design choices that can achieve wide dynamic range are adjusting the amplification rate or using high resolution digital converters. However, the former requires sophisticated, power-hungry noise reduction circuits to achieve the desirable dynamic range while the latter incurs expensive, high power consumption convertors.
Although this technique eliminates the need for a high-precision ADC or an adjustable amplifier, Nemo's processor must quickly react to changes in current and adjust the resistor value.
Depending on the input current (which is supported in the range 1$\mu$A and 200mA), the resolution of Nemo varies in the range 0.013$\mu$A to 48$\mu$A.
%The resolution of Nemo is the difference between input current of two adjacent ADC readings.
% ******** The measurement error of Nemo is 8.3\% when compared to the current reported by a DMM.
% Especially, the maximum error is observed for very low current values.
Unfortunately, Nemo does not measure bus voltage, and the evaluations are simple and do not include an IoT device with high power variations.
%Therefore, when the voltage variation range is high, the shunt switching delay and ADC turn off time may significantly affect the power measurement accuracy.
%----
% The maximum error, 8.3%, occurs at the lowest end of the dynamic range. When the input current is larger than 10 uA, most of the errors fall below 2%.
%evaluations used TelosB
%Quote: We have implemented a prototype of Nemo (as shown in Fig. 2) and conducted extensive experimental evaluation. Our results show that Nemo has satisfactory measurement fidelity under a range of operating conditions. In particular, Nemo yields a dynamic measurement range from 0.8 uA to 200 mA, a sampling rate of 8 KHz, and a minimum resolution of 0.013 uA, while only incurring an average measurement error of 1.34%.
% a table compares iCount and SPOT - 
%Quote: Since we do not have access to the iCount and SPOT hardware, the performance data of the two systems are ob- tained from two papers [5] [11]. Tab. 2 summarizes the comparison of the three power metering systems.
% THERE IS A VERY NICE COMPARISON TABLE IN THE PAPER
% -----------

%--------------------------------
$\mu$Monitor \cite{Naderiparizi2016} proposes a power monitoring platform based on counting capacitor charging and discharge cycles. %which is similar to Energy Bucket \cite{Andersen2009}.
%The energy overhead of $\mu$Monitor is 1.7 $\mu$A for operating voltage 3.5 V.
For the loads within the range of 1$\mu$W to 10mW, the accuracy of $\mu$Monitor is almost within 10\% of the results obtained from a 16-bit ADC that digitizes the voltage value over a shunt resistor.
Unfortunately, the evaluations use static loads.
% Additionally, the authors did not discuss about the sampling rate of this platform.
%Quote: Using a Keysight N6705B Power Analyzer (with built-in ampere- and voltmeter) as ref- erence system a Matlab script calculates a regression line for each analog input parameter and stores offset and slope to an EEPROM
% -------------------------------
%2017
Similar to Nemo \cite{Zhou2013a}, Potsch et al. \cite{Potsch2017,Potsch2014} propose the use of two shunt resistors ($1\Omega$ and $100\Omega$) for measuring low and high currents up to 100mA.
%The second resistor can be bypassed by a MOSFET switch.
Shunt resistor voltages are amplified and then sampled by a 16-bit ADC. %(AD7682).
A 32-bit microcontroller communicates with the ADC and collects the samples.
% Therefore, although the ADC supports 15Ksps, the microcontroller may not be able to handle this rate.
The actual sampling rate and resolution of this platform have not been evaluated.
%NICE: the paper provides an overview of various commercial power measurement systems
% we compare the accuracy of these platforms against a DMM
% this enables us to only measure current - can we measure voltage as well?
% run the raspberry pi programs in user and kernel spaces and measure the energy consumption of the baords

In addition to the limitations highlighted, a common shortcoming of the platforms reviewed in this section is their complex circuitry.
Specifically, these platforms include various components such as ADCs, op-amps, a resistor series, high precision capacitors, and a processor.
Therefore, building these platforms is costly and time consuming.
In contrast, EMPIOT is very easy to build, and the cost of a complete platform (shield plus base board) is around \$35 when an RPi3 is used as the base board and \$15 when an RPiZW is used.

\subsubsection{Limited Range}
%--------------------------------
%--------------------------------
PowerBench \cite{Haratcherev2008} is capable of providing a 5KHz sampling rate and 30$\mu$A resolution.
%As the systems uses a 12-bit ADC, the system resolution is 16 $\mu$A, without taking into account the noise error of the ADC.
% They claim that, as an effective carrier sense using CC1000 takes about 0.5 ms, a sampling rate of 5 KHz would be enough to capture all events on a Mica2 board.
PowerBench assumes that the maximum current consumption of the host device is 65mA.
This platform samples the amplified voltage (51x) across a shunt resistor using a 12-bit ADC.
%PowerBench has been used in a wireless sensor network testbed to evaluate the accuracy of estimation-based energy measurement.
% We replaced the normal TNOde with a resistor to precisely control the current to be measured. That current was measured simul- taneously by a precise (≤1%) multimeter and our power trace capturing setup. We recorded the difference and incorporated that into a calibration table (one entry per node) that assures compensation for both absolute and proportional errors. (This table is used by the runner when processing the raw ADC values.) By using resistors with different values the whole current measurement range (0-65mA) could be calibrated.
%--------------------------------
%--------------------------------
% [-] requiring building a circuit
The authors in \cite{Trathnigg2008} assume energy profiling requires current measurement only.
The error of this platform is less than 5\% as long as current is higher than 20$\mu$A.
However, the maximum supported current is 35mA. 
%Using this platform, the authors proposes enhancements for the accuracy of PowerTOSSIM \cite{VictorShnayder}, which is a power estimation simulator.
%--------------------------------
% [-] no clear evaluation
%Zhu and O'Connor \cite{Zhu2013b} highlighted the importance of energy measurement for high data rate and low-power devices.
%They choose nRF24L01, which is a low-power and high data rate (2 Mbps) device.
iWEEP\_HW \cite{Zhu2013b} employs a multi-layer architecture to measure the power consumption of processor, transceiver, and sensors using separate ADC channels.
This platform uses a PIC18 processor and a 10-bit ADC which measures voltage across shunt resistors.
iWEEP\_HW can measure currents up to 40mA with a maximum sampling rate of 150KHz.
% *****Unfortunately, the measurement error of this platform is unclear.
%NICE: they have explored the energy consummation of the processor in various states and also when running different instructions
%As shown in Table II, SPI and UART communications in MCU do not consume extra energy. 
%So do the instruction executions of MCU, the instructions under investigation are AND, OR, XOR, ADD, DIV, MUL, WHILE and FOR loop [6]
%Using this platform, the authors measured the energy consumed by the processor and transceiver when sending packets.
%----------------------
PotatoScope \cite{Hartung2016} is a microcontroller-based oscilloscope that focuses on reliable energy measurement in outdoor environments and in the presence of significant temperature variation.
PotatoScope includes an ARM Cortex-M3 processor that is attached to a 12-bit ADC to sample current and voltage.
%The ADC supports sampling rates up to 500 KHz and its accuracy is 5 LSB.
The voltage across a 0.47$\Omega$ shunt resistor is amplified by 200x.
Since the ADC's reference voltage is 2.5V, the maximum measurable current is 26.6mA.
% The resolution of PotatoScope is 0.917mV for voltage and 6.5$\mu$A for current.
%PotatoScope relies on the integrated Direct Memory Access (DMA) Controller of the processor to transfer the collected data to the SD card.
% NOTE: The paper shows a mathematical model for error computation - nice!
% We applied a voltage to the circuit and observed a remark- able ADC dispersion. Results show that ADC values vary within a range of up to 60 LSB. This random noise is caused by several sources, e.g. lab conditions, power supply or other components of the PotatoScope itself. However, this means that values scatter ± 30 LSB around the expected value which corresponds to an error of 0.7% when considering a resolution of 12 bit. From this it follows that we cannot reduce errors below 0.7% due to noise in the signal chain.
% Instead of the SSH tunnel, a more lightweight TCP connection is used to reduce the transmission delays between PotatoScope and central server.
%NICE: the have used a signal generator for their evaluations

The main shortcoming of these platforms is their limited current measurement range, which is less than 100mA.
In this paper we showed that the new generation of IoT devices rely on high data rate technologies that increase temporal power consumption as high as 700mA.
Therefore, none of the platforms studied in this section can be used with these IoT devices.
The performance evaluation results presented in Section \ref{perf_eval} confirm the effectiveness of EMPIOT for measuring the energy of 802.11 IoT devices.

\subsubsection{Low Accuracy}

%%---------
% [-] requiring building a circuit

Energino \cite{Gomez2012a} uses the 10-bit ADC of Arduino boards.
%Quote: it is mandatory to collect samples with a very fine granularity, ideally in the order of 10 mW or less
%The Arduino board provides six input channels connected to a 10-bit ADC.
The ADC supports an input voltage of 0 to 5V, therefore, the LSB is 4.88mV and the current measurement resolution is 25mA.
Energino uses a Hall-effect current sensor with sensitivity 185mV/A to measure currents up to 5A.
% Although the theoretical sampling rate is 10kHz, the sampling rate is 1KHz.
The actual sampling rate is 1KHz.
%Quote: Such a sampling rate is not high enough to catch MAC-level events such as acknowledgments however, we accepted it as a reasonable trade–off between cost/complexity and performance.
%The Arduino board has been extended with three components: voltage divider, current sensor, and a relay.
%In order to measure input voltages as high as 55V, a voltage divider is used to map the input voltage to range 0 to 5V.
%The third component is a relay, which is used for turning on/off the device.
%The Arduino board continuously samples the voltage and current sensors, accumulates them, and sends a report over the USB interface.
%The platform is controlled through a management back-end developed by Python.
%%---------
%%---------
%%---------
%%---------
NITOS \cite{Keranidis2014b} uses the ATmega2560 micro-controller with a 10-bit ADC.
% This micro-controller provides 10-bit ADC channels, 256 KB flash memory and 8 KB SRAM.
%In order to support remote control and access to the measurement platform, an Ethernet Arduino shield with a SD card module has been added to the design.
Since the Arduino's ADC cannot measure millivolt level variations in voltage, a voltage amplifier has been used.
%The amplification of INA139 depends on the values of the resistors attached to this chip.
%Furthermore, the ADC reference voltage has been set to 2.56V.
%The authors assume that a typical MPDU would take 27$\mu$s (= 1534B/450Mbps); therefore, the 9 KHz sampling rate of Arduino would not be enough to capture these behavior.
In order to maximize sampling rate, NITOS: (i) uses ADC free running mode, (ii) increases the ADC's prescalar clock from 125KHz to 1MHz, and (iii) uses interrupt service routine to get ADC values.
Although these enhancements result in up to a 63KHz sampling rate, the accuracy of ADC is reduced by 11\% due to the higher clock used.
Furthermore, the current measurement resolution of the platform is 25mA.

{\color{black!50!black}
\subsection{Analytical Estimation}
\label{related_ana}
In addition to empirical measurement, analytical modeling can be employed to estimate power.
In \cite{wang2006realistic} the authors model the energy consumption of wireless communications for single and multi-hop networks.
However, since the authors rely on a simple CSMA protocol, the model would not be valid in the presence of interference and collision.
Furthermore, the energy consumption of the processor and other components such as sensors, has not been taken into account.
Similar to this work, unfortunately, most of the existing models only consider the energy consumption of wireless communication \cite{du2010modeling,li2007analytical}.
In particular, due to its prevalence, 802.15.4 is the primary technology modeled.

In addition to communication aspects, models have been proposed for modeling the energy consumption of processors \cite{bazzaz2013accurate,konstantakos2008energy}.
After accurate energy profiling of processor and peripherals, the energy cost of each instruction is used to calculate the energy of applications.

The authors in \cite{martinez2015power} propose a framework for modeling the power consumption of various components on an IoT device.
To model the power consumption of networking, they take into account the effect of transmission power, re-transmission, and spreading factor, as used by wireless technologies such as LoRa.
Processing power has been modeled based on the average power consumed per arithmetic instruction and the time complexity of the algorithm used.
To apply the models to a system, however, it is essential to first run experiments and fit the models into the empirical results collected.

Although analytical approaches are very useful during the system design phase, empirical energy measurement is inevitable.
In addition to the points we mentioned in Section \ref{intro}, the followings points further justify the importance of empirical evaluation.
First, accurate and thorough energy profiling is necessary to formulate analytical energy estimation models.
For example, the energy consumption of the board might be significantly affected by environmental interference or the start-up power of components.
Second, empirical long term energy analysis is necessary to identify the hidden sources of energy depletion \cite{brusey2016energy}.
For example, some studies show the significant effect of link breakage and server outage on energy usage.
Third, various types of batteries are affected differently by load and environmental factors.
}

{\color{black!50!black}
\section{Conclusion}
\label{conclusion}
The EMPIOT platform presented in this paper enables low-cost, programmable, and accurate energy measurement of a wide range of IoT devices.
After explaining the hardware and software components of this platform, we evaluated the effect of various parameters on performance.
Our studies show that: 
(i) the BCM driver results in a higher sampling rate and lower energy consumption, 
(ii) lowering the shield's operational voltage slightly reduces the sampling rate, 
(iii) voltage variations do not affect accuracy,
and (iv) the effect of two-buffer or circular buffering mechanisms on power consumption of the base board depends on the number of processor cores.
In terms of calibration, we developed a programmable load that enables us to generate a wide range of load values and calibrate the platform without requiring precise timing control.

For IoT devices whose minimum energy consumption is more than 100$\mu$A, EMPIOT can be simply connected to the input power supply of the device.
If the minimum energy consumption is less than 100$\mu$A, EMPIOT uses a hybrid energy estimation model to take into account the time spent in sleep modes.
To this end, extra connections are required to inform EMPIOT before and after every sleep duration.
We evaluated the performance of EMPIOT using five different IoT devices and four types of workloads.
Our results confirm that the energy measurement error is less than 3.5\% for IoT devices that utilize either 802.15.4 or 802.11 wireless standards. 
Beyond the scope of this paper, our study on the I2C performance and energy efficiency of Linux provides insights into designing Linux-based IoT systems.

Some of the future work avenues are as follows: 
Although the energy consumption of simple wireless technologies such as 802.15.4 and LoRa have been thoroughly studied and modeled, in addition to 802.11b/n standards, newer technologies, such as 802.11ac, NB-IoT, and eMTC, are being used for IoT applications. 
The EMPIOT platform provides a low-cost and scalable solution to deploy testbeds and profile the energy consumption of these complex wireless technologies.
Another area of future work is to port EMPIOT to non-Linux-based systems and measure its sampling rate and overhead when the operating system of the base board is a RTOS. Not only limited to IoT devices, the measurement range of EMPIOT makes it a suitable energy measurement platform for other applications, such as measuring and evaluating the effect of energy efficiency techniques proposed for cloud computing platforms \cite{aldawsari2015trusted,baker2017energy}.
}

\section*{Acknowledgment}
This work has been partially supported by a research grant from Cypress Semiconductor Corporation (Grant No. CYP001).

% or


% \appendices

% % you can choose not to have a title for an appendix
% % if you want by leaving the argument blank
% \section{}

% \section*{Acknowledgment}

\ifCLASSOPTIONcaptionsoff
  \newpage
\fi



%\bibliographystyle{IEEEtran}
\bibliographystyle{IEEEtran}
\bibliography{bibliography}

\begin{thebibliography}{10}
\providecommand{\url}[1]{#1}
\csname url@samestyle\endcsname
\providecommand{\newblock}{\relax}
\providecommand{\bibinfo}[2]{#2}
\providecommand{\BIBentrySTDinterwordspacing}{\spaceskip=0pt\relax}
\providecommand{\BIBentryALTinterwordstretchfactor}{4}
\providecommand{\BIBentryALTinterwordspacing}{\spaceskip=\fontdimen2\font plus
\BIBentryALTinterwordstretchfactor\fontdimen3\font minus
  \fontdimen4\font\relax}
\providecommand{\BIBforeignlanguage}[2]{{%
\expandafter\ifx\csname l@#1\endcsname\relax
\typeout{** WARNING: IEEEtran.bst: No hyphenation pattern has been}%
\typeout{** loaded for the language `#1'. Using the pattern for}%
\typeout{** the default language instead.}%
\else
\language=\csname l@#1\endcsname
\fi
#2}}
\providecommand{\BIBdecl}{\relax}
\BIBdecl

\bibitem{Gartner_IoT}
\BIBentryALTinterwordspacing
{\relax Gartner}. \relax {IoT Units Installed Base by Category (Millions of
  Units)}. [Online]. Available:
  \url{https://www.gartner.com/newsroom/id/3598917}
\BIBentrySTDinterwordspacing

\bibitem{Kaup2014}
F.~Kaup, P.~Gottschling, and D.~Hausheer, ``{PowerPi: Measuring and modeling
  the power consumption of the Raspberry Pi},'' in \emph{Proceedings of the
  39th Annual IEEE Conference on Local Computer Networks (LCN'14)}, 2014, pp.
  236--243.

\bibitem{Gomez2012a}
K.~Gomez, R.~Riggio, T.~Rasheed, D.~Miorandi, and F.~Granelli, ``{Energino: a
  Hardware and Software Solution for Energy Consumption Monitoring},'' in
  \emph{Proceedings of the International Workshop on Wireless Network
  Measurements (WiOpt'12)}, 2012, pp. 311 -- 317.

\bibitem{Zhou2013a}
R.~Zhou and G.~Xing, ``{Nemo: A high-fidelity noninvasive power meter system
  for wireless sensor networks},'' \emph{Proceedings of the ACM/IEEE
  International Conference on Information Processing in Sensor Networks
  (IPSN'13)}, pp. 141--152, 2013.

\bibitem{Trathnigg2008}
T.~Trathnigg, M.~J{\"u}rgen, and R.~Weiss, ``A low-cost energy measurement
  setup and improving the accuracy of energy simulators for wireless sensor
  networks,'' in \emph{Proceedings of the workshop on Real-world wireless
  sensor networks}, 2008, pp. 31--35.

\bibitem{eriksson2009accurate}
J.~Eriksson, F.~{\"O}sterlind, N.~Finne, A.~Dunkels, N.~Tsiftes, and T.~Voigt,
  ``Accurate network-scale power profiling for sensor network simulators.'' in
  \emph{Proceedings of the International Conference on Embedded Wireless
  Systems and Networks (EWSN'09)}, vol.~9.\hskip 1em plus 0.5em minus
  0.4em\relax Springer, 2009, pp. 312--326.

\bibitem{wang2006realistic}
Q.~Wang, M.~Hempstead, and W.~Yang, ``A realistic power consumption model for
  wireless sensor network devices,'' in \emph{Annual IEEE Communications
  Society on Sensor and Ad Hoc Communications and Networks (SECON)},
  vol.~1.\hskip 1em plus 0.5em minus 0.4em\relax IEEE, 2006, pp. 286--295.

\bibitem{martinez2015power}
B.~Martinez, M.~Monton, I.~Vilajosana, and J.~D. Prades, ``The power of models:
  Modeling power consumption for iot devices,'' \emph{IEEE Sensors Journal},
  vol.~15, no.~10, pp. 5777--5789, 2015.

\bibitem{Li2013b}
Q.~Li, M.~Martins, O.~Gnawali, and R.~Fonseca, ``{On the effectiveness of
  energy metering on every node},'' in \emph{Proceedings of the IEEE
  International Conference on Distributed Computing in Sensor Systems
  (DCoSS'13)}, 2013, pp. 231--240.

\bibitem{dezfouli2015modeling}
B.~Dezfouli, M.~Radi, S.~A. Razak, T.~Hwee-Pink, and K.~A. Bakar, ``Modeling
  low-power wireless communications,'' \emph{Journal of Network and Computer
  Applications}, vol.~51, pp. 102--126, 2015.

\bibitem{dezfouli2014cama}
B.~Dezfouli, M.~Radi, K.~Whitehouse, S.~A. Razak, and H.-P. Tan, ``Cama:
  Efficient modeling of the capture effect for low-power wireless networks,''
  \emph{ACM Transactions on Sensor Networks (TOSN)}, vol.~11, no.~1, p.~20,
  2014.

\bibitem{CYW43907AEVAL}
\BIBentryALTinterwordspacing
{\relax Cypress Semiconductor}. {\relax CYW943907AEVAL1F Evaluation Kit}.
  [Online]. Available:
  \url{http://www.cypress.com/documentation/development-kitsboards/cyw943907aeval1f-evaluation-kit}
\BIBentrySTDinterwordspacing

\bibitem{CC2650}
\BIBentryALTinterwordspacing
{\relax CC2650 SimpleLink Multistandard Wireless MCU}. [Online]. Available:
  \url{https://www.ti.com/lit/ds/swrs158b/swrs158b.pdf}
\BIBentrySTDinterwordspacing

\bibitem{Stathopoulos2008}
T.~Stathopoulos, D.~McIntire, and W.~J. Kaiser, ``{The energy endoscope:
  Real-time detailed energy accounting for wireless sensor nodes},''
  \emph{Proceedings of International Conference on Information Processing in
  Sensor Networks (IPSN'08)}, pp. 383--394, 2008.

\bibitem{brusey2016energy}
J.~Brusey, J.~Kemp, E.~Gaura, R.~Wilkins, and M.~Allen, ``Energy profiling in
  practical sensor networks: Identifying hidden consumers,'' \emph{IEEE Sensors
  Journal}, vol.~16, no.~15, pp. 6072--6080, 2016.

\bibitem{34465A}
\BIBentryALTinterwordspacing
{\relax Keysight Technologies}. \relax {34465A (6$\frac{1}{2}$ digit)}.
  [Online]. Available:
  \url{https://literature.cdn.keysight.com/litweb/pdf/5991-1983EN.pdf}
\BIBentrySTDinterwordspacing

\bibitem{kth_dmm_7510}
\BIBentryALTinterwordspacing
{\relax Tektronix}. \relax {DMM7510 7$\frac{1}{2}$ Digit Graphical Sampling
  Multimeter}. [Online]. Available:
  \url{https://www.tek.com/tektronix-and-keithley-digital-multimeter/dmm7510}
\BIBentrySTDinterwordspacing

\bibitem{Monsoon}
\BIBentryALTinterwordspacing
{\relax Monsoon Inc.} \relax {Monsoon Power Monitor}. [Online]. Available:
  \url{https://www.msoon.com/LabEquipment/PowerMonitor/}
\BIBentrySTDinterwordspacing

\bibitem{manweiler2011avoiding}
J.~Manweiler and R.~Roy~Choudhury, ``Avoiding the rush hours: Wifi energy
  management via traffic isolation,'' in \emph{Proceedings of the 9th
  international conference on Mobile systems, applications, and
  services}.\hskip 1em plus 0.5em minus 0.4em\relax ACM, 2011, pp. 253--266.

\bibitem{serrano2015per}
P.~Serrano, A.~Garcia-Saavedra, G.~Bianchi, A.~Banchs, and A.~Azcorra,
  ``Per-frame energy consumption in 802.11 devices and its implication on
  modeling and design,'' \emph{IEEE/ACM Transactions on Networking (ToN)},
  vol.~23, no.~4, pp. 1243--1256, 2015.

\bibitem{Jiang2007}
X.~Jiang, P.~Dutta, D.~Culler, and I.~Stoica, ``{Micro Power Meter for Energy
  Monitoring of Wireless Sensor Networks at Scale},'' \emph{Proceedings of the
  6th international conference on Information processing in sensor networks
  (IPSN'07)}, p. 186, 2007.

\bibitem{Duttac2008}
P.~Dutta, M.~Feldmeier, J.~Paradiso, and D.~Culler, ``Energy metering for free:
  Augmenting switching regulators for real-time monitoring,'' in
  \emph{Proceedings of the 7th International Conference on Information
  Processing in Sensor Networks (IPSN'08)}, 2008, pp. 283--294.

\bibitem{Andersen2009}
J.~Andersen and M.~T. Hansen, ``{Energy Bucket: A Tool for Power Profiling and
  Debugging of Sensor Nodes},'' in \emph{Proceedings of Third International
  Conference on Sensor Technologies and Applications (SENSORCOMM'09)}.\hskip
  1em plus 0.5em minus 0.4em\relax IEEE, 2009, pp. 132--138.

\bibitem{Naderiparizi2016}
S.~Naderiparizi, A.~N. Parks, F.~S. Parizi, and J.~R. Smith, ``{$\mu$Monitor:
  In-situ energy monitoring with microwatt power consumption},'' in
  \emph{Proceedings of the IEEE International Conference on RFID
  (RFID'16)}.\hskip 1em plus 0.5em minus 0.4em\relax IEEE, may 2016, pp. 1--8.

\bibitem{Haratcherev2008}
I.~Haratcherev, G.~Halkes, and T.~Parker, ``{PowerBench: A Scalable Testbed
  Infrastructure for Benchmarking Power Consumption},'' in \emph{Proceedings of
  the International Workshop on Sensor Network Engineering (IWSNE'08)}, 2008,
  pp. 37--44.

\bibitem{Zhu2013b}
N.~Zhu and I.~O'Connor, ``Energy measurements and evaluations on high data rate
  and ultra low power wsn node,'' in \emph{10th IEEE International Conference
  on Networking, Sensing and Control (ICNSC)}, 2013, pp. 232--236.

\bibitem{Hartung2016}
R.~Hartung, U.~Kulau, and L.~Wolf, ``{Distributed energy measurement in WSNs
  for outdoor applications},'' \emph{Proceedings of the 13th Annual IEEE
  International Conference on Sensing, Communication, and Networking
  (SECON'16)}, 2016.

\bibitem{Potsch2017}
A.~P{\"o}tsch, A.~Berger, and A.~Springer, ``Efficient analysis of power
  consumption behaviour of embedded wireless iot systems,'' in
  \emph{Proceedings of the Instrumentation and Measurement Technology
  Conference (I2MTC)}, 2017, pp. 1--6.

\bibitem{Potsch2014}
A.~P{\"{o}}tsch, A.~Berger, C.~Leitner, and A.~Springer, ``{A power measurement
  system for accurate energy profiling of embedded wireless systems},'' in
  \emph{Proceedings of the 19th IEEE International Conference on Emerging
  Technologies and Factory Automation (ETFA'14)}, 2014, pp. 1--4.

\bibitem{Keranidis2014b}
S.~Keranidis, G.~Kazdaridis, V.~Passas, T.~Korakis, I.~Koutsopoulos, and
  L.~Tassiulas, ``{NITOS Energy Monitoring Framework: Real Time Power
  Monitoring in Experimental Wireless Network Deployments},'' \emph{SIGMOBILE
  Mob. Comput. Commun. Rev.}, vol.~18, no.~1, pp. 64--74, 2014.

\bibitem{INA219}
\BIBentryALTinterwordspacing
\relax{INA219 Zero-Drift, Bidirectional Current/Power Monitor With I2C
  Interface}. [Online]. Available:
  \url{http://www.ti.com/lit/ds/symlink/ina219.pdf}
\BIBentrySTDinterwordspacing

\bibitem{bcmdriver}
\BIBentryALTinterwordspacing
``\relax{C library for Broadcom BCM 2835}.'' [Online]. Available:
  \url{http://www.airspayce.com/mikem/bcm2835/}
\BIBentrySTDinterwordspacing

\bibitem{linuxi2c}
\BIBentryALTinterwordspacing
``\relax{Linux I2C Driver}.'' [Online]. Available:
  \url{https://www.kernel.org/doc/Documentation/i2c/dev-interface}
\BIBentrySTDinterwordspacing

\bibitem{ina219adafruit}
\BIBentryALTinterwordspacing
``\relax{Adafruit INA219 Current Sensor Breakout}.'' [Online]. Available:
  \url{https://learn.adafruit.com/adafruit-ina219-current-sensor-breakout/downloads}
\BIBentrySTDinterwordspacing

\bibitem{strace}
\BIBentryALTinterwordspacing
\relax{strace: Linux syscall tracer}. [Online]. Available:
  \url{https://strace.io}
\BIBentrySTDinterwordspacing

\bibitem{CYW43907}
\BIBentryALTinterwordspacing
{\relax Cypress Semiconductor}. {\relax CYW43907: IEEE 802.11 a/b/g/n SoC with
  an Embedded Applications Processor}. [Online]. Available:
  \url{http://www.cypress.com/file/298236/download}
\BIBentrySTDinterwordspacing

\bibitem{U8001}
\BIBentryALTinterwordspacing
{\relax Keysight Technologies.} \relax {U8000 Series, Single Output DC Power
  Supplies}. [Online]. Available:
  \url{https://literature.cdn.keysight.com/litweb/pdf/5989-7182EN.pdf?id=1460471}
\BIBentrySTDinterwordspacing

\bibitem{Lim2013a}
R.~Lim, F.~Ferrari, M.~Zimmerling, C.~Walser, P.~Sommer, and J.~Beutel,
  ``{FlockLab: A Testbed for Distributed, Synchronized Tracing and Profiling of
  Wireless Embedded Systems},'' in \emph{Proceedings of the 12th international
  conference on Information processing in sensor networks (IPSN'13)}, 2013, p.
  153.

\bibitem{AD5200}
\BIBentryALTinterwordspacing
{\relax Analog Devices}. \relax{256-Position and 33-Position Digital
  Potentiometers}. [Online]. Available:
  \url{http://www.analog.com/media/en/technical-documentation/data-sheets/AD5200\_5201.pdf}
\BIBentrySTDinterwordspacing

\bibitem{ADG1612}
\BIBentryALTinterwordspacing
------. \relax{Quad SPST Switches}. [Online]. Available:
  \url{http://www.analog.com/media/en/technical-documentation/data-sheets/ADG1611\_1612\_1613.pdf}
\BIBentrySTDinterwordspacing

\bibitem{TI_RTOS}
\BIBentryALTinterwordspacing
\relax{TI-RTOS: Real-Time Operating System (RTOS)}. [Online]. Available:
  \url{http://www.ti.com/tool/TI-RTOS}
\BIBentrySTDinterwordspacing

\bibitem{BCM4343W}
\BIBentryALTinterwordspacing
{\relax BCM4343W: 802.11b/g/n WLAN, Bluetooth and BLE SoC Module}. [Online].
  Available:
  \url{https://products.avnet.com/opasdata/d120001/medias/docus/138/AES-BCM4343W-M1-G\_data\_sheet\_v2\_3.pdf}
\BIBentrySTDinterwordspacing

\bibitem{WICED}
\BIBentryALTinterwordspacing
``{\relax WICED: Wireless Internet Connectivity for Embedded Devices)}.''
  [Online]. Available: \url{http://www.cypress.com/products/wiced-software}
\BIBentrySTDinterwordspacing

\bibitem{feeney2001investigating}
L.~M. Feeney and M.~Nilsson, ``Investigating the energy consumption of a
  wireless network interface in an ad hoc networking environment,'' in
  \emph{Twentieth Annual Joint Conference of the IEEE Computer and
  Communications Societies (INFOCOM'01)}, 2001, pp. 1548--1557.

\bibitem{Milenkovic2005}
A.~Milenkovic, M.~Milenkovic, E.~Jovanov, D.~Hite, and D.~Raskovic, ``{An
  environment for runtime power monitoring of wireless sensor network
  platforms},'' in \emph{Proceedings of the Thirty-Seventh Southeastern
  Symposium on System Theory (SSST'05)}, 2005, pp. 406--410.

\bibitem{MCC}
\BIBentryALTinterwordspacing
{\relax Measurement Computing Corporation}. {\relax USB-1608FS-Plus data
  acquisition device}. [Online]. Available:
  \url{http://www.mccdaq.com/pdfs/manuals/USB-1608FS-Plus.pdf}
\BIBentrySTDinterwordspacing

\bibitem{ti_bat_mon}
\BIBentryALTinterwordspacing
{\relax BQ2019: Advanced Battery Monitor}. [Online]. Available:
  \url{https://www.ti.com/lit/ds/slus465e/slus465e.pdf}
\BIBentrySTDinterwordspacing

\bibitem{du2010modeling}
W.~Du, F.~Mieyeville, and D.~Navarro, ``Modeling energy consumption of wireless
  sensor networks by systemc,'' in \emph{Fifth International Conference on
  Systems and Networks Communications (ICSNC)}.\hskip 1em plus 0.5em minus
  0.4em\relax IEEE, 2010, pp. 94--98.

\bibitem{li2007analytical}
J.~Li and P.~Mohapatra, ``Analytical modeling and mitigation techniques for the
  energy hole problem in sensor networks,'' \emph{Pervasive and Mobile
  Computing}, vol.~3, no.~3, pp. 233--254, 2007.

\bibitem{bazzaz2013accurate}
M.~Bazzaz, M.~Salehi, and A.~Ejlali, ``An accurate instruction-level energy
  estimation model and tool for embedded systems,'' \emph{IEEE transactions on
  instrumentation and measurement}, vol.~62, no.~7, pp. 1927--1934, 2013.

\bibitem{konstantakos2008energy}
V.~Konstantakos, A.~Chatzigeorgiou, S.~Nikolaidis, and T.~Laopoulos, ``Energy
  consumption estimation in embedded systems,'' \emph{IEEE Transactions on
  instrumentation and measurement}, vol.~57, no.~4, pp. 797--804, 2008.

\bibitem{aldawsari2015trusted}
B.~Aldawsari, T.~Baker, and D.~England, ``Trusted energy efficient cloud-based
  services brokerage platform,'' \emph{Int. J. Intell. Comput. Res}, vol.~6,
  pp. 630--639, 2015.

\bibitem{baker2017energy}
T.~Baker, M.~Asim, H.~Tawfik, B.~Aldawsari, and R.~Buyya, ``An energy-aware
  service composition algorithm for multiple cloud-based iot applications,''
  \emph{Journal of Network and Computer Applications}, vol.~89, pp. 96--108,
  2017.

\end{thebibliography}



%\vfill


\end{document}